\documentclass[sn-mathphys,Numbered]{sn-jnl}

\usepackage{graphicx}%
\usepackage{multirow}%
\usepackage{amsmath,amssymb,amsfonts}%
\usepackage{amsthm}%
\usepackage{mathrsfs}%
\usepackage[title]{appendix}%
\usepackage{xcolor}%
\usepackage{textcomp}%
\usepackage{manyfoot}%
\usepackage{booktabs}%
\usepackage{algorithm}%
\usepackage{algorithmicx}%
\usepackage{algpseudocode}%
\usepackage{listings}%

\definecolor{dkgreen}{rgb}{0,0.6,0}
\definecolor{gray}{rgb}{0.5,0.5,0.5}
\definecolor{mauve}{rgb}{0.58,0,0.82}

\theoremstyle{thmstyleone}%
\theoremstyle{thmstyletwo}%

\theoremstyle{thmstylethree}%

\begin{document}

\title[Article Title]{Stochastic Differential Equation for a System Coupled to a Thermostatic Bath via an Arbitrary Interaction Hamiltonian}


\author[1,2,3]{\fnm{Jong-Min} \sur{Park}}\email{jongmin.park@apctp.org}

\author[4]{\fnm{Hyunggyu} \sur{Park}}\email{hgpark@kias.re.kr}

\author*[3]{\fnm{Jae Sung} \sur{Lee}}\email{jslee@kias.re.kr}

\affil[1]{\orgname{Asia Pacific Center for Theoretical Physics}, \orgaddress{\city{Pohang}, \postcode{37673}, \country{Korea}}}

\affil[2]{\orgdiv{Department of Physics}, \orgname{Postech}, \orgaddress{\city{Pohang}, \postcode{37673}, \country{Korea}}}

\affil[3]{\orgdiv{School of Physics}, \orgname{Korea Institute for Advanced Study}, \orgaddress{\city{Seoul}, \postcode{02455}, \country{Korea}}}

\affil[4]{\orgdiv{Quantum Universe Center}, \orgname{Korea Institute for Advanced Study}, \orgaddress{\city{02455}, \postcode{610101}, \country{Korea}}}


\abstract{The conventional Langevin equation offers a mathematically convenient framework for investigating open stochastic systems interacting with their environment or a bath. However, it is not suitable for a wide variety of systems whose dynamics rely on the nature of the environmental interaction, as the equation does not incorporate any specific information regarding that interaction. Here, we present a universal formulation of the stochastic differential equation (SDE) for an open system coupled to a thermostatic bath via an arbitrary interaction Hamiltonian. This SDE encodes the interaction information to a fictitious potential (mean force) and a position-dependent damping coefficient. Surprisingly, we find that the conventional Langevin equation can be recovered in the presence of arbitrary strong interactions given two conditions: translational invariance of the potential and disjoint separability of the bath particles. Our results provide a universal framework for studying open stochastic systems with an arbitrary interaction Hamiltonian and yield deeper insight into why various experiments fit the conventional Langevin description regardless of the strength or type of interaction. 
}

\maketitle

\section{Introduction}\label{sec1}

An open system interacts with its environment or bath, and its dynamics depend on the nature of the system–bath (SB) interaction. Therefore, information on the full Hamiltonian, including the environment, is necessary to understand and describe the dynamics of an open system accurately. In most cases, however, the huge number of degrees of freedom for the environment renders this approach impractical. Thus, an effective and phenomenological equation of motion is required to describe the process of an open system approximately. In this context, a mesoscopic stochastic differential equation (SDE), such as the Langevin equation, has been widely used to successfully describe open systems with continuous time and space. This SDE, where simple dissipation and noise terms are used to represent the SB interaction effect, has enabled the derivation of important relations in stochastic thermodynamics, including fluctuation theorems~\cite{Jarzynski1997nonequilibrium, Crooks1999entropy, Kurchan1998, Seifert2005ep}, thermodynamic uncertainty relations~\cite{Barato2015thermodynamics, Gingrich2016dissipation, Gingrich2017inferring, Pietzonka2017finite, Horowitz2017proof, Dechant2019multidimensional, Hasegawa2019uncertainty, Vu2019uncertainty, Lee2019thermodynamic}, and speed limit~\cite{shiraishi2018speed, van2020unified, van2021geometrical, yoshimura2021thermodynamic, ito2020stochastic, falasco2020dissipation, nicholson2020time, tasnim2021thermodynamic, dechant2022minimum, delvenne2021thermo,lee2022speed}.

Although the conventional SDE is successful, it fails to capture the specific nature of the SB interaction. This interaction can be neglected at the macroscopic scale, given that the SB interaction Hamiltonian is inversely proportional to the system size in comparison to the volume Hamiltonian. However, at a small scale, the SB interaction Hamiltonian is comparable to the system Hamiltonian and thus cannot be neglected, as it will affect the dynamics of the system significantly~\cite{kwon2019three,strasberg2017stochastic}. Unfortunately, the specific nature of the SB interaction is not included in the conventional Langevin equation, limiting the ability to study small open systems that are strongly coupled to a bath~\cite{seifert2016first,jarzynski2017stochastic}. As a result, the thermodynamics of small systems with strong SB interaction has not been investigated systematically.

In this study, we develop an SDE to capture the nature of the SB interactions, which is applicable to a system coupled to a thermostatic bath via arbitrary SB interactions. The proposed SDE encodes information regarding the SB interaction into two terms: the \emph{mean-force} term, which is a fictitious potential force applied to a system that modifies the equilibrium state from the ordinary Gibbs state of the system Hamiltonian, and a position-dependent damping tensor. Remarkably, we identify two physical conditions that can lead to the vanishing of SB interaction effects, even in the case of strong coupling. These conditions correspond to the \emph{translational invariance} of the interaction potentials and the \emph{disjoint separability} of the bath particles. Under these conditions, we demonstrate that our SDE reduces to the conventional Langevin equation without incorporating any information regarding the SB interaction specifics. This clearly shows that the conventional Langevin equation and the Gibbs state of the system Hamiltonian can be obtained as an equilibrium state without the assumption of a weak interaction.

\section{SDE for Arbitrary System-Bath Interactions}\label{sec2}

\subsection{Microscopic Setup}\label{subsec2_1}

\begin{figure}
	\centering
	\includegraphics[width = \columnwidth]{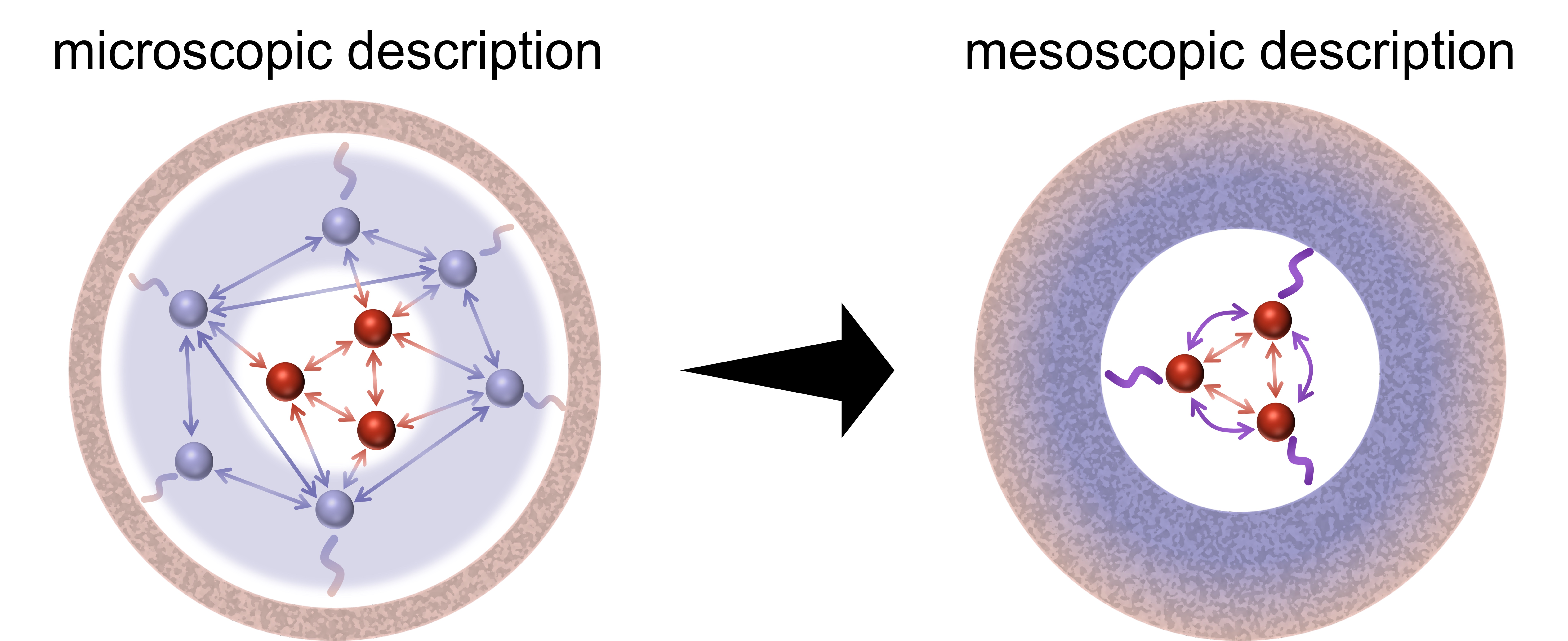}
	\caption{Schematic illustration of the microscopic description (a) and the effective mesoscopic description (b). Red and blue balls denote the system and bath particles, respectively, and the outer reddish ring represents the thermostat connected to bath particles in (a).  Wavy curves and arrows   depict thermal contact to thermostat and the interactions between particles, respectively. In (b), the outer circular area with gradient colors and the wavy purple curves denote the effective bath and the thermal contact between system particles and the effective bath, respectively.  }
	\label{fig:model}
\end{figure}

We begin by considering an SB composite, as illustrated in Fig.~\ref{fig:model}(a). The system and the bath consist of $N$ and $\tilde{N}$ particles, each with mass $m$ and $\tilde m$, respectively. The particles interact with each other via certain potentials. We will use the notation $\tilde{\scriptstyle{\square}}$ for denoting quantities belonging to the bath. For brevity of presentation, we restrict our discussion to a one-dimensional space;
however, extension to higher dimensional spaces is straightforward. The position and velocity of the $n$th system particle at time $t$ are denoted as $x_n(t)$ and $v_n(t)$ for $n=1,2,...,N$, respectively. Similarly, $\tilde{x}_{\tilde n} (t)$ and $\tilde{v}_{\tilde{n}} (t)$ for $\tilde{n} = 1,2,...,\tilde{N}$ denote the position and velocity of the $\tilde n$th bath particle at time $t$. For maintaining the bath temperature, the bath particles are connected to a Langevin thermostat with temperature $T$ and damping constant $\tilde \gamma$. This is a typical setup for various molecular dynamics (MD) simulations. The microscopic equations of motion for the system and bath particles are given by
\begin{align}
	{\rm system}:&~\boldsymbol{v} (t)= \dot{\boldsymbol{x}} (t),~~m \dot{\boldsymbol{v}} (t)
	= \boldsymbol{f} ( \boldsymbol{x}(t), \boldsymbol{v}(t), t )
	- \boldsymbol{\nabla}_x
	V_\textrm{I} (\boldsymbol{x}(t), \tilde{\boldsymbol{x}} (t)), \nonumber \\
	{\rm bath}:&~\tilde{\boldsymbol{v}} (t)= \dot{\tilde{ \boldsymbol{x}}} (t),~~ \tilde{m} \dot{\tilde{\boldsymbol{v}}} (t)
	= - \boldsymbol{\nabla}_{\tilde{x}}
	V_\textrm{I} (\boldsymbol{x}(t), \tilde{\boldsymbol{x}} (t)) - {\tilde \gamma} \tilde{\boldsymbol{v}} (t)
	+ \tilde{\boldsymbol{\xi}} (t), \label{eq:micro_bath}
\end{align}
where $\boldsymbol{x} = (x_1, \cdots, x_N)^{ \textsf T}$, $\boldsymbol{v} = (v_1,\cdots, v_N)^{\textsf T}$, $\tilde{\boldsymbol{x}} = (\tilde{x}_1, \cdots, \tilde{x}_{\tilde{N}})^{\textsf T}$, $\tilde{\boldsymbol{v}} = (\tilde{v}_1, \cdots, \tilde{v}_{\tilde{N}})^{\textsf T}$, $\boldsymbol{\nabla}_x = (\partial_{x_1}, \cdots,\partial_{x_N})^{\textsf T}$, and $\boldsymbol{\nabla}_{\tilde{x}} = (\partial_{\tilde{x}_1}, \cdots, \partial_{\tilde{x}_{\tilde N}})^{\textsf T}$, with $\textsf T$ denoting the matrix transpose. $\boldsymbol{f}(\boldsymbol{x},\boldsymbol{v},t)$ represents a force acting on the system particles and $V_\textrm{I} (\boldsymbol{x},\tilde{\boldsymbol{x}}) = H_\textrm{I} (\boldsymbol{x},\tilde{\boldsymbol{x}}) + \tilde{\Phi} (\tilde{\boldsymbol{x}})$ is associated with the interaction potential, where $H_\textrm{I} (\boldsymbol{x},\tilde{\boldsymbol{x}})$ is the SB interaction Hamiltonian and $\tilde{\Phi} (\tilde{\boldsymbol{x}})$ is the potential exerted only on the bath particles. $\tilde{\boldsymbol{\xi}} = (\tilde \xi_1, \cdots, \tilde \xi_{\tilde N})^{\textsf T}$ is the Gaussian white noise with zero mean and $\langle \tilde{\xi}_i (t) \tilde{\xi}_j (t') \rangle = 2 {\tilde \gamma} T \delta_{ij} \delta (t-t')$ in Boltzmann constant units ($k_{\rm B}=1$). Note that the thermostat connected to the bath has been referred to as a \emph{superbath} in the literature~\cite{Jarzynski1997nonequilibrium,strasberg2017stochastic,kwon2019three}.

Equilibrium of the total system (system and bath) is achievable when a conservative force is solely applied to the system, that is, $\boldsymbol{f}(\boldsymbol{x},\boldsymbol{v},t) = - \boldsymbol{\nabla}_x \Phi (\boldsymbol{x})$. Then, the equilibrium distribution $p^\textrm{eq}$ of the total system is simply given by the Gibbs state as
\begin{equation}
	p^\textrm{eq} (\boldsymbol{x},\boldsymbol{v},\tilde{\boldsymbol{x}},\tilde{\boldsymbol{v}})
	\equiv
	\frac{1}{\mathcal{Z}}
	e^{-\beta ( H (\boldsymbol{x},\boldsymbol{v}) + H_\textrm{I} (\boldsymbol{x},\tilde{\boldsymbol{x}}) + \tilde{H} (\tilde{\boldsymbol{x}},\tilde{\boldsymbol{v}}) )}, \label{eq:GibbsDistTot}
\end{equation}
where $H (\boldsymbol{x},\boldsymbol{v}) = m \boldsymbol{v}^{\textsf T}\boldsymbol{v} /2 + \Phi (\boldsymbol{x})$ is the system Hamiltonian, $\tilde{H} (\tilde{\boldsymbol{x}}, \tilde{\boldsymbol{v}}) = \tilde{m} \tilde{\boldsymbol{v}}^{\textsf T} \tilde{\boldsymbol{v}} /2 + \tilde{\Phi} (\tilde{\boldsymbol{x}})$ is the bath Hamiltonian, $\mathcal{Z}$ is the partition function for the total Hamiltonian $H+H_{\rm I}+\tilde H$, and $\beta = 1/T$ is the inverse temperature. Integrating out Eq.~\eqref{eq:GibbsDistTot} over the bath variables yields the equilibrium distribution for the system as follows:
\begin{equation}
	p_{\rm sys}^\textrm{eq} (\boldsymbol{x},\boldsymbol{v}) \equiv \int d\tilde{\boldsymbol{x}} d\tilde{\boldsymbol{v}} ~p^\textrm{eq} (\boldsymbol{x},\boldsymbol{v},\tilde{\boldsymbol{x}},\tilde{\boldsymbol{v}})
	= \frac{1}{\mathcal {Z}_{\rm sys} }
	e^{-\beta (H(\boldsymbol{x},\boldsymbol{v}) + \Delta(\boldsymbol{x}))}, \label{eq:eq_dist_sys}
\end{equation}
where $\mathcal {Z}_{\rm sys}$ is the partition function for the Hamiltonian $H + \Delta$ and $\Delta (\boldsymbol{x})$ is a fictitious potential known as the potential of the mean force~\cite{seifert2016first}, defined as
\begin{equation}\label{def:mean_force_potential}
	\Delta (\boldsymbol{x}) \equiv - T
	\ln
	\frac{\mathcal{Z}_{\rm I}(\boldsymbol{x})}{\mathcal Z_{\tilde \Phi}} 
\end{equation}
with $\mathcal{Z}_{\rm I} = \int d\tilde{\boldsymbol{x}} e^{-\beta V_I (\boldsymbol{x}, \tilde{\boldsymbol{x}})}$ and $\mathcal{Z}_{\tilde \Phi} = \int d\tilde{\boldsymbol{x}} e^{-\beta \tilde{\Phi} (\tilde{\boldsymbol{x}})}$. The distribution $p_{\rm sys}^{\rm eq}$ in Eq.~\eqref{eq:eq_dist_sys} is certainly different from the Gibbs state of the system Hamiltonian $p_{\rm sys}^{\rm G} \equiv e^{-\beta H}/{\mathcal{Z}_{\rm sys}^{\rm G}}$ with $\mathcal Z_{\rm sys}^{\rm G} = \int d \boldsymbol{x} d \boldsymbol{v} e^{-\beta H (\boldsymbol{x}, \boldsymbol{v})}$, unless $\Delta = 0$. Since vanishing of $\Delta$ is achieved for negligible $H_{\rm I}$, it is generally accepted that the weak interaction (small $H_{\rm I}$) limit is necessary for an equilibrium distribution of a system being $p_{\rm sys}^{\rm G}$. However, since the weak interaction limit amounts to isolating the system from the environment, heat transfer or relaxation is unlikely to occur under these conditions~\cite{talkner2020colloquium,van1957approach}. This contradicts the usual Langevin-system experiments, in which relaxation takes place quickly and the equilibrium state is given by $p_{\rm sys}^{\rm G}$. This strongly suggests that there exists another mechanism that leads to $\Delta$ vanishing, rather than the weak interaction.

\subsection{Brief Sketch of Mesoscopic SDE Derivation}

The derivation of the mesoscopic SDE for a system is based on time-scale separation, in which the variables of the bath are treated as fast variables compared to those of the system. The derivation procedure can be divided into two steps. The first step is taking the overdamped limit of the equation of motion for the bath, which is mathematically equivalent to the limit of small $\tilde m/\tilde \gamma$. The second step is called \emph{adiabatic elimination}~\cite{risken1996fokker}, where the bath variables are integrated out in the small $\tilde \gamma$ limit. The details are presented in the Appendix. After integrating out all bath variables via this procedure, the resulting mesoscopic SDE is 
\begin{align}\label{eq:reduced_Langevin}
	m \dot{\boldsymbol{v}} (t) =
	\boldsymbol{f} (\boldsymbol{x}(t), \boldsymbol{v}(t), t)
	- \boldsymbol{\nabla}_x \Delta(\boldsymbol{x}(t)) - \mathsf{G} (\boldsymbol{x}(t)) \cdot \boldsymbol{v} (t)
	+ \boldsymbol{\xi}(t)~,   
\end{align}
where the Gaussian white noise $\boldsymbol{\xi} = ( \xi_1, \cdots, \xi_{ N})^{\textsf T}$ is characterized by $\langle \boldsymbol{\xi} (t) \boldsymbol{\xi}^{\textsf T} (t') \rangle = 2 T \mathsf{G}(\boldsymbol{x}(t)) \delta(t - t')$ with zero mean. Here, $\mathsf{G} (\boldsymbol{x})$ is the effective damping tensor given by
\begin{equation}\label{def:damping_tensor}
	\mathsf G_{n,m}(\boldsymbol{x}) = \frac{1}{T}
	\int_0^\infty dt~
	C_{\partial_{x_n} V_\textrm{I}, \partial_{x_m} V_\textrm{I}} (t|\boldsymbol{x})~, 
\end{equation}
with the correlation $C_{h,g} (t|\boldsymbol{x})$ of two arbitrary functions $h(\boldsymbol{x},\tilde{\boldsymbol{x}})$ and $g(\boldsymbol{x},\tilde{\boldsymbol{x}})$ defined as
\begin{equation}\label{def:correl_func}
	C_{h,g} (t|\boldsymbol{x}) \equiv
	\left \langle
	h\left( \boldsymbol{x},\tilde{\boldsymbol{x}}(t) \right)
	g \left( \boldsymbol{x},\tilde{\boldsymbol{x}}(0) \right) 
	\right \rangle_{\rm b}^{\rm eq}
	- \left \langle h\left( \boldsymbol{x},\tilde{\boldsymbol{x}} \right) \right \rangle_{\rm b}^{\rm eq}
	\left \langle g \left( \boldsymbol{x},\tilde{\boldsymbol{x}} \right) \right \rangle_{\rm b}^{\rm eq},
\end{equation}
where $\langle \cdots \rangle_{\rm b}^{\rm eq}$ denotes the ensemble average with respect to the equilibrium distribution of the bath formed via Eq.~\eqref{eq:micro_bath} for the given particle position $\boldsymbol{x}$.
Note that the damping tensor $\mathsf{G} (\boldsymbol{x})$ in Eq.~\eqref{def:damping_tensor} depends on the specific form of the SB interaction $H_\textrm{I} (\boldsymbol{x},\tilde{\boldsymbol{x}})$ and is symmetric since $C_{f,g} (t|\boldsymbol{x}) = C_{g,f} (t|\boldsymbol{x})$, guaranteed by the micro-reversibility of the equilibrium
bath.
We further note that a similar damping tensor has recently been found~\cite{ding2023multi} through the Nakajima–Zwanzig projection operator method~\cite{nakajima1958quantum,zwanzig1960ensemble} for deterministic Hamiltonian dynamics.

The SDE~\eqref{eq:reduced_Langevin} includes two terms reflecting the strong coupling effects: (i) the potential of the mean force $\Delta$ and (ii) the effective damping tensor $\mathsf G$. Since the fluctuation–dissipation relation holds, $\mathsf G$ has no effect on the equilibrium state, but only affects the relaxation dynamics. On the other hand, $\Delta$ modifies the equilibrium distribution in accordance with Eq.~\eqref{eq:eq_dist_sys}. Thus, this SDE provides a universal mesoscopic description, as illustrated in Fig.~\ref{fig:model}(b), for studying dynamics, as well as the steady states of open stochastic systems with arbitrary interaction Hamiltonians.

\section{Conditions for Vanishing of Interaction Specifics} \label{sec:two_conditions}

\begin{figure}
	\centering
	\includegraphics[width = \columnwidth]{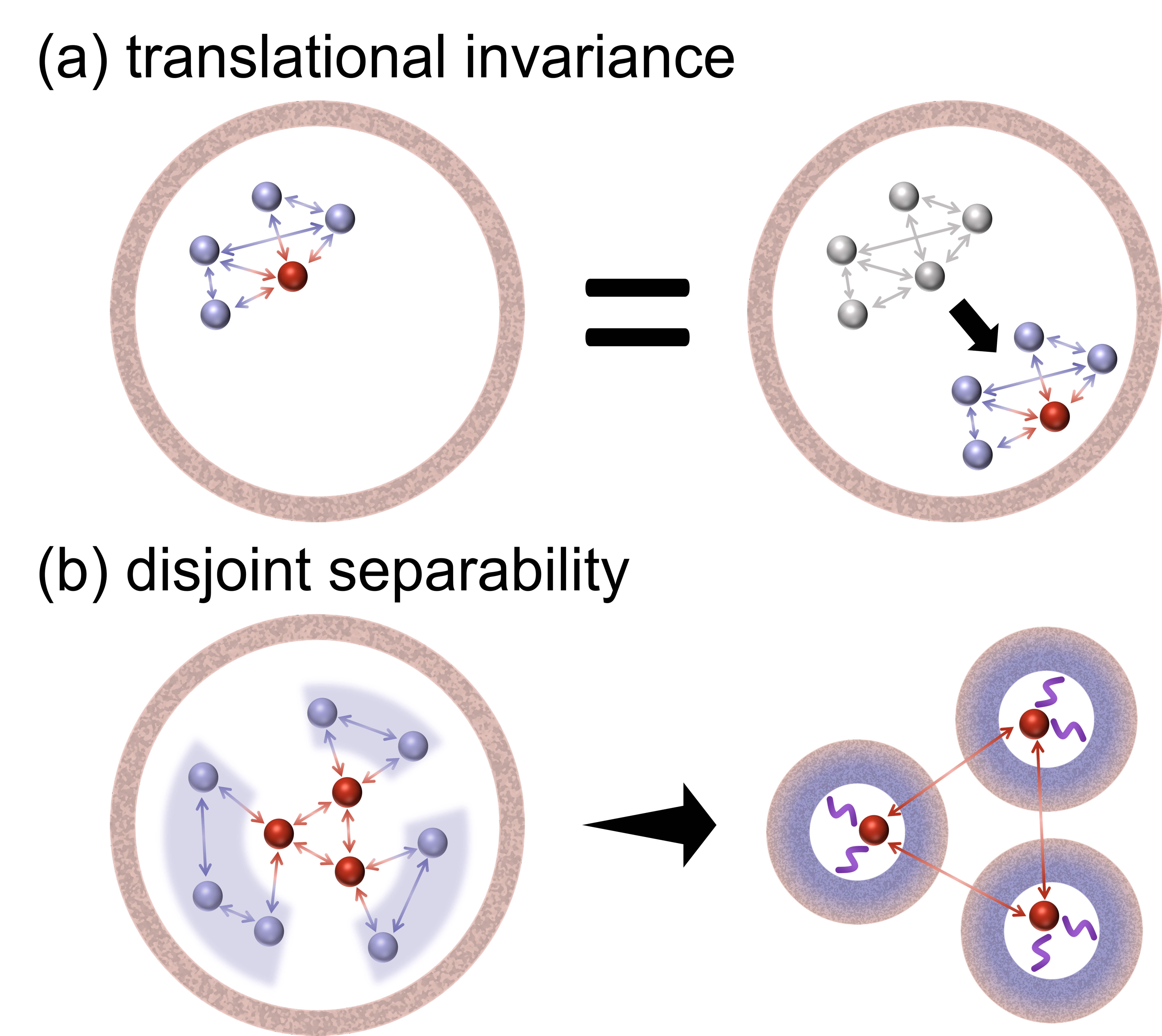}
	\caption{
		Schematic diagram of the two conditions for vanishing of interaction specifics. (a) Translational invariance. This is the condition that interaction potentials for any pairs of particles are invariant under an arbitrary translational shift. (b) Disjoint separability. This condition disallows that each bath particle interacts with multiple system particles simultaneously. Thus, all bath particles can be disjointly partitioned into subgroups interacting with respective system particles. Interactions between bath particles belonging to other groups are prohibited. 
	}
	\label{fig:conditions}
\end{figure}

We find that two physical conditions are necessary for the interaction specifics to vanish and to return Eq.~\eqref{eq:reduced_Langevin} to the conventional Langevin equation. The first is \emph{translational invariance} of the interaction potential $V_\textrm{I} (\boldsymbol{x}, \tilde{\boldsymbol{x}} )$ in space, meaning $V_\textrm{I} (\boldsymbol{x}, \tilde{\boldsymbol{x}} ) = V_\textrm{I} (\boldsymbol{x} + \boldsymbol{a}, \tilde{\boldsymbol{x}} + \tilde{\boldsymbol{a}} )$, where $\boldsymbol{a}$ and $\tilde{\boldsymbol{a}}$ are $N$- and $\tilde N$-dimensional vectors, respectively, with all elements being equal to an arbitrary constant $a$ in one space dimension.
Under this condition, the interaction potential for any pair of particles is not altered by an arbitrary translational shift, as illustrated in Fig.~\ref{fig:conditions}(a). This condition leads to the translational invariance of the environmental influence, as follows:
\begin{equation}\label{eq:conditionI}
	\Delta (\boldsymbol{x} + \boldsymbol{a})
	= \Delta (\boldsymbol{x})
	~~\textrm{and}~~
	\mathsf{G} (\boldsymbol{x} + \boldsymbol{a})
	= \mathsf{G} (\boldsymbol{x}),
\end{equation}
which can be verified using Eqs.~\eqref{def:mean_force_potential}, \eqref{def:damping_tensor} and the translational invariance of the correlation $C_{\partial_{x_n} V_\textrm{I}, \partial_{x_m} V_\textrm{I}} (t|\boldsymbol{x} + \boldsymbol{a}) = C_{\partial_{x_n} V_\textrm{I}, \partial_{x_m} V_\textrm{I}} (t|\boldsymbol{x})$. The condition of translational invariance is usually valid for experiments implemented in the bulk region (far from the boundary) of their environment.
Note that the force acting on the system 
$f(\boldsymbol{x},\boldsymbol{v}, t)$ may not satisfy
the translational invariance.

The second is the \emph{disjoint separability} of the bath particles. This condition prohibits each bath particle from interacting with multiple (more than one) system particles simultaneously. Under this condition, one can disjointly partition all bath particles into subgroups interacting with respective system particles, as depicted in Fig.~\ref{fig:conditions}(b). The interaction between bath particles belonging to other groups is also prohibited. Disjoint separability is approximately valid when the typical length scale between system particles is much larger than the range of the SB interaction. The effect of the SB interaction range on disjoint separability is numerically tested in Sec.~\ref{sec:separability}.

With these two conditions, first we can show that the mean-force term  in Eq.~\eqref{eq:reduced_Langevin} vanishes.
The translational invariance of $V_{\rm I}$ ensures that its form should be
expressed in a pairwise manner as explained in Appendix ~\ref{sec:app4}.
Then, incorporating the disjoint separability together, $V_{\rm I}$ can be written as 
\begin{equation}
	H_\textrm{I} (\boldsymbol{x},\tilde{\boldsymbol{x}})
	= \sum_{n,\tilde{n}_n} H_{n, \tilde{n}_n} (\tilde{x}_{\tilde{n}_n} - x_n )
	~~\textrm{and}~~
	\tilde{\Phi} (\tilde{\boldsymbol{x}}) = \sum_{n, \tilde{n}_n > \tilde{m}_n}
	\tilde{\Phi}_{\tilde{n}_n,\tilde{m}_n} (\tilde{x}_{\tilde{n}_n} - \tilde{x}_{\tilde{m}_n}), \label{eq:pairwise_interaction}
\end{equation}
where $\tilde n_n$ is index for bath particles belonging to a subgroup interacting with the $n$th system particle. By changing the variable as $\tilde X_{\tilde n_n} = \tilde x_{\tilde n_n} - x_n $, the potential of the mean force~\eqref{def:mean_force_potential} is expressed as 
\begin{equation}
	\Delta  = - T \ln \int d\tilde{\boldsymbol{X}}~ e^{-\beta \sum_{n,\tilde{n}_n} H_{n, \tilde{n}_n} ( \tilde{X}_{\tilde{n}_n})   -\beta \sum_{n, \tilde{n}_n > \tilde{m}_n}
		\tilde{\Phi}_{\tilde{n}_n,\tilde{m}_n} (\tilde{X}_{\tilde{n}_n} - \tilde{X}_{\tilde{m}_n})  }
	+ T \ln \mathcal{Z}_{\tilde \Phi }, \label{eq:vanishing_mean_force_pot}
\end{equation}
where $\int d\tilde{\boldsymbol{X}}$ indicates integration over all $\tilde X_{\tilde n_n}$ variables. 
Since the integrand in Eq.~\eqref{eq:vanishing_mean_force_pot}  is a function of only $\tilde X_{\tilde n_n}$, $\Delta$ has no dependence of position variables of the system. Therefore, the mean force term $\boldsymbol{\nabla}_x \Delta $ in Eq.~\eqref{eq:reduced_Langevin} vanishes. 

We can also show that dependence of the damping tensor on the SB interaction potential disappears. Disjoint separability leads to the following force balance relation for the subgroup of bath particles interacting with $n$th system particle:
\begin{equation} 
	-\partial_{x_n} V_\textrm{I} (\boldsymbol{x}, \tilde{\boldsymbol{x}})
	- \sum_{\tilde{n}_n} \partial_{\tilde{x}_{\tilde{n}_n}}
	V_\textrm{I} (\boldsymbol{x}, \tilde{\boldsymbol{x}}) = 0. \label{eq:balance_relation}
\end{equation}
Using Eq.~\eqref{eq:balance_relation}, we rewrite  Eq.~\eqref{def:damping_tensor} as
\begin{equation}\label{def:simple_damping_tensor}
	\mathsf G_{n,m} = \frac{1}{T} \sum_{\tilde n_n, \tilde n_m}
	\int_0^\infty dt~
	C_{\partial_{\tilde x_{\tilde n_n}} V_\textrm{I}, \partial_{\tilde x_{\tilde n_m}} V_\textrm{I}} (t|\boldsymbol{x}).  
\end{equation}
The correlation $C$ in Eq.~\eqref{def:simple_damping_tensor} is written in terms of the force applied to a bath particle while the same $C$ in Eq.~\eqref{def:damping_tensor} is expressed by the force exerted on a system particle. Therefore, we can now apply the generalized Green-Kubo relation to Eq.~\eqref{def:simple_damping_tensor} as presented in Appendix \ref{sec:app2}, which yields
\begin{equation}\label{eq:GK_rel}
	\int_0^\infty dt
	C_{\partial_{\tilde{x}_{\tilde{n}_n}} V_\textrm{I},
		\partial_{\tilde{x}_{\tilde{n}_m}} V_\textrm{I}} (t|\boldsymbol{x})
	=  \tilde \gamma T \delta_{\tilde{n}_n,\tilde{n}_m}. 
\end{equation}
Therefore, the damping tensor is simplified as 
\begin{equation}\label{eq:restored_G}
	\textsf G_{n,m}  = \gamma_n  \delta_{n,m},~~~~ (\gamma_n \equiv \tilde{N}_n \tilde{\gamma}) 
\end{equation}
where $\tilde{N}_n$ is the number of bath particles belonging to the subgroup interacting with $n$th system particle. Consequently, dependence of $\textsf G$ on the SB interaction potential and $\boldsymbol{x}$ completely disappears. 

Therefore, under the two conditions, Eq.~\eqref{eq:reduced_Langevin} is reduced to the conventional Langevin equation as
\begin{equation}\label{eq:restored_Langevin}
m \dot{v}_n (t) =
f_n (\boldsymbol{x}(t), \boldsymbol{v}(t), t) - \gamma_n v_n(t) + \xi_n (t), 
\end{equation}
where $\langle \xi_n(t) \xi_m (t^\prime) \rangle = 2 \gamma_n T \delta_{nm} \delta(t-t^\prime) $. 
This clearly shows that the conventional Langevin equation can be derived even in the presence of strong coupling. It is also worthwhile to note that disjoint separability is automatically satisfied for a single-particle system. Thus, the conventional Langevin description is valid for a single-particle system if only translational invariance is satisfied.

\section{Numerical Confirmation}
To verify our analytic results, we performed numerical simulations for two solvable examples with harmonic couplings. The first example is a single particle coupled to a confined bath without translational invariance and the second one is a two-particle system coupled to a bath without disjoint separability. In addition, we carried out MD simulations where two-system particles interact with bath particles
through a soft elastic repulsion proportional to their overlapped length.

\subsection{Single-particle model without translational invariance} \label{sec:example1}

\begin{figure}
	\centering
	\includegraphics[width = \columnwidth]{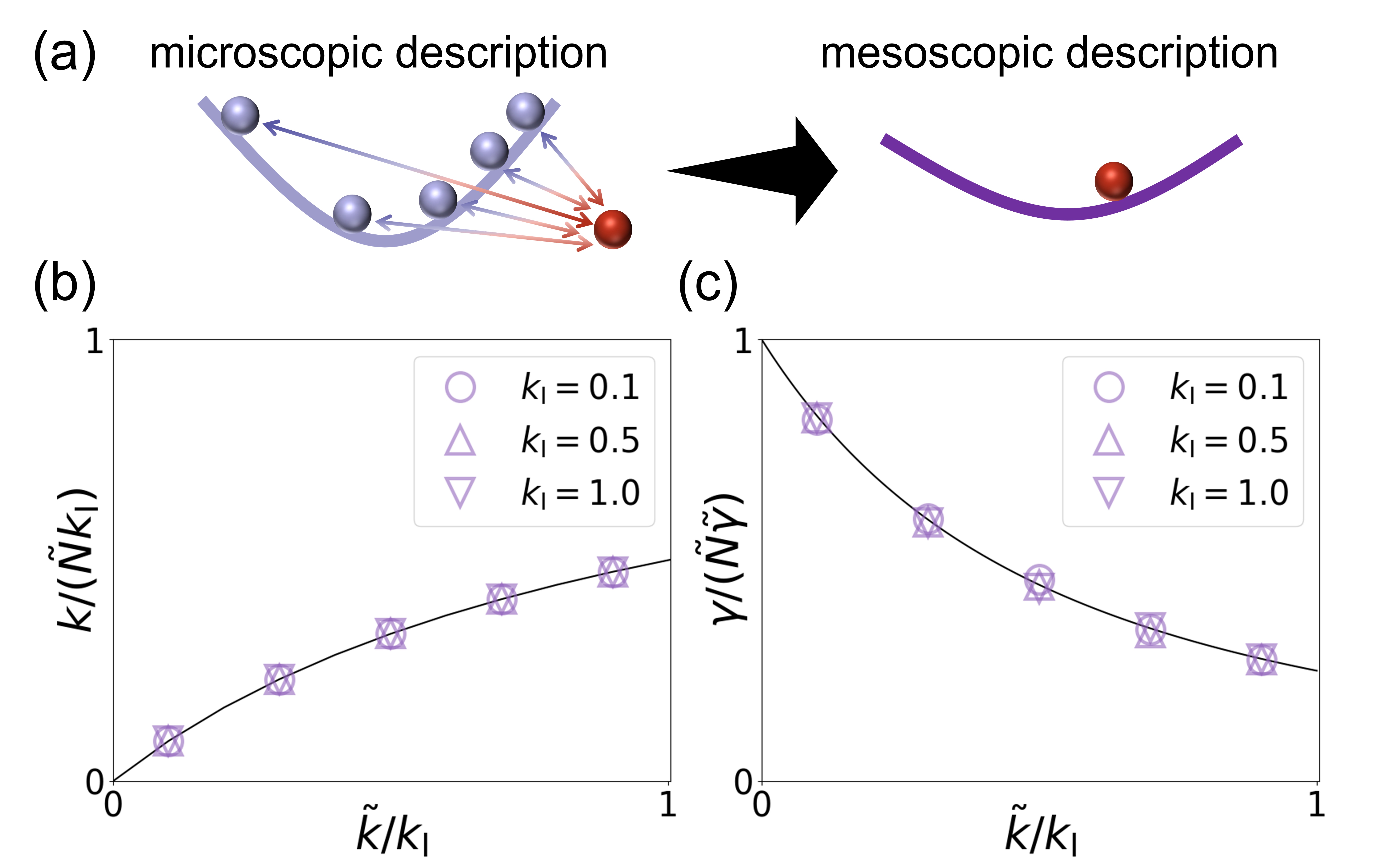}
	\caption{Simulation results of the single-particle model without translation invariance. (a) Schematic of the model. Interactions between the system and bath particles in the microscopic description can be converted into the effective harmonic potential in the mesoscopic description. (b) Plot for $k/(\tilde N k_{\rm I})$ versus $\tilde{k}/k_{\rm I}$ for various $k_{\rm I}$. (c) Plot for $\gamma/(\tilde N \tilde \gamma)$ versus $\tilde k / k_{\rm I}$ for various $k_{\rm I}$.  The black curves denote the analytical predictions. 
	}
	\label{fig:ex1}
\end{figure}

Here, we consider a single-particle system as illustrated in Fig.~\ref{fig:ex1}(a). $\tilde N$ bath particles ($\tilde{\boldsymbol{x}}$) are confined in a harmonic potential centered at the origin with stiffness $\tilde k$ and the system particle ($x_1$) is coupled to each bath particle via a harmonic potential with stiffness $k_{\rm I}$. Thus, the interaction potential $V_{\rm I} = H_{\rm I} + \tilde \Phi$ is given by
\begin{equation}
     H_{\rm I} (x_1,\tilde{\boldsymbol{x}}) = 
     \sum_{\tilde{n}}  \frac{1}{2} k_{\rm I}  ( x_1 - \tilde{x}_{\tilde{n}}  )^2,~~
     \tilde \Phi (\tilde{\boldsymbol{x}}) = 
     \frac{1}{2} \tilde{k} \tilde{\boldsymbol{x}}^{\rm T} \tilde{\boldsymbol{x}}~. 
\end{equation}
The confining potential $\tilde \Phi (\tilde{\boldsymbol{x}} )$ breaks the translational invariance of $V_{\rm I}$. On the other hand, the disjoint separability is satisfied since the system consists of a single particle. 

For evaluating the potential of the mean force, it is convenient to rearrange $V_{\rm I}$ as 
\begin{equation}
	V_{\rm I} (x_1, \tilde{\boldsymbol{x}}) =
	\sum_{\tilde{n}}
	\frac{1}{2} \left( k_{\rm I} + \tilde{k} \right) \tilde X_{\tilde n}^2
	+ \frac{\tilde{N} k_{\rm I} \tilde{k}}{2 ( k_{\rm I} + \tilde{k} ) }
	x_1^2, 
\end{equation}
where $\tilde X_{\tilde n} = \tilde{x}_{\tilde{n}} –  k_{\rm I} x_1 /( k_{\rm I} + \tilde{k} ) $. Then, from Eq.~\eqref{def:mean_force_potential}, the potential of the mean force is 
\begin{equation}
\Delta (x_1) = \frac{1}{2} k x_1^2 +
\frac{\tilde N T}{2} \ln
\frac{k_{\rm I} + \tilde{k}}{\tilde{k}}, \label{eq:meanforce_ex1}
\end{equation}
where $k \equiv \tilde{N} k_{\rm I} \tilde{k} / ( k_{\rm I} + \tilde{k} )$ is the effective stiffness applied to the system particle. This $\Delta (x_1)$ is tantamount to the harmonic force, $-kx_1$, being exerted on the system particle, which is the consequence of the confinement or broken translational symmetry of the bath particles. It is straightforward to see that the mean force vanishes when the translational symmetry is restored, i.e. $\tilde{k}=0$.

The damping constant~\eqref{def:damping_tensor} is also analytically solvable. Using the Hermitianized Fokker-Plank operator presented in Appendix~\ref{sec:app3}, the effective damping constant ($\gamma \equiv \textsf G_{1,1}$) can be evaluated as
\begin{equation}\label{eq:damping_ex1}
\gamma = \tilde{N} \tilde{\gamma}
\left (\frac{k_{\rm I}}{k_{\rm I} + \tilde{k}}
\right )^2~.
\end{equation}
Thus, dependence of $\gamma$ on the interaction strength $k_{\rm I}$ disappears in the limit $\tilde{k} \rightarrow 0$ irrespective of the value of $k_{\rm I}$.  This is consistent with the analytic prediction~\eqref{eq:restored_G} that the damping constant becomes independent of interaction potential when the translational invariance is satisfied for a one-particle system. 

Equations~\eqref{eq:meanforce_ex1} and \eqref{eq:damping_ex1} indicate that dynamics of the system can be effectively described by the mesoscopic SDE as
\begin{equation}
	m \dot{v}_1 (t) = f(x_1(t),v_1(t),t) -k x_1 (t) - \gamma v_1 (t) + \sqrt{2 \gamma T} \xi_1 (t).
\end{equation}
Note that the effective damping constant $\gamma$ vanishes in the weak-interaction limit ($k_I \rightarrow 0$), which implies the isolation of the system from the thermal environment. This clearly demonstrates that the conventional
Langevin description is not the consequence of the weak interaction.

To confirm our analytic results numerically, we performed a simulation
using the microscopic equations of motion \eqref{eq:micro_bath} with small $\tilde{m}/\tilde{\gamma}$ and small $\tilde{\gamma}$ for proper time-scale separation.

For simplicity, no external force is applied to the system, i.e., $f(x,v,t)=0$. In this calculation $\tilde{N} = 10^4$, $\tilde{\gamma}=10^{-2}$, $\tilde{m} = 10^{-4}$, and $m=T=1$ were used. We evaluated $k$ and $\gamma$
by means of measuring the variance of the position distribution and evaluating the Green-Kubo formula in equilibrium, respectively, for various values of $k_{\rm I}$ and $\tilde{k}$. To reduce the computational cost, we used the equation of motion for the center of mass coordinate of the bath particles as explained in  Appendix.  Figures~\ref{fig:ex1}(b) and (c) are the plots for $k/(\tilde N k_{\rm I})$ versus $\tilde k / k_{\rm I}$ and $\gamma / (\tilde N \tilde \gamma)$ versus $\tilde k/k_{\rm I}$, respectively. Respective numerical data perfectly fit the analytic formulae of the effective stiffness in Eq.~\eqref{eq:meanforce_ex1} and the effective damping constant~\eqref{eq:damping_ex1}. This confirms that for a one-particle system the strong coupling effect appears when the translational invariance of the interaction potential  is broken.

\subsection{Two-particle model without disjoint separability} \label{sec:example2}

\begin{figure}
	\centering
	\includegraphics[width = \columnwidth]{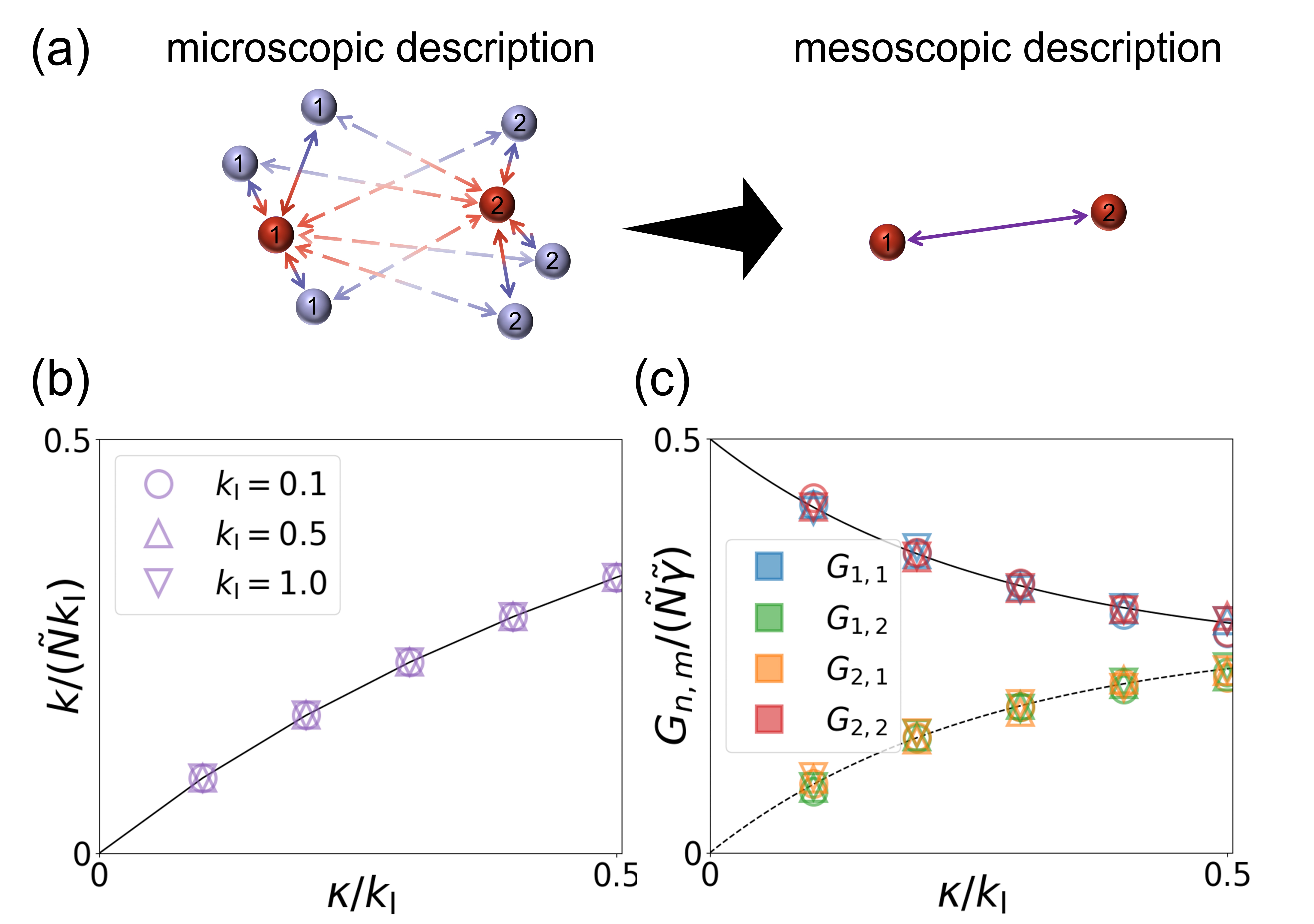}
	\caption{Simulation results of the two-particle model without disjoint separability. (a) Schematic of the model. The solid and dashed arrows indicate the intra-interactions inside the same group and inter-interaction between other group particles, respectively. The system-bath Interactions in the microscopic description can be converted into the effective harmonic potential in the mesoscopic description. (b) Plot for $k/(\tilde N k_{\rm I})$ versus $\kappa /k_{\rm I}$ for various $k_{\rm I}$. The black solid curve denotes the analytical prediction. (c) Plot for $\textsf G_{n,m}/(\tilde N \tilde \gamma)$ versus $\kappa / k_{\rm I}$ for various $k_{\rm I}$.  The black solid and dashed curves denote the analytical predictions for diagonal and off-diagonal elements of $\textsf G$. Circle, up-pointing triangle, and down-pointing triangle represent the numerical data for $k_{\rm I} = 0.1, 0.5$, and  $1.0$, respectively. Blue, green, orange, and red colors in (c) represent the data for $\textsf G_{1,1}$, $\textsf G_{1,2}$, $\textsf G_{2,1}$ and $\textsf G_{2,2}$, respectively.
	}
	\label{fig:ex2}
\end{figure}

The second example is a two-particle system interacting with bath particles of two different species as illustrated in Fig.~\ref{fig:ex2}(a). In this model, each bath particle is coupled to both system particles via a harmonic potential simultaneously, but with different stiffness depending on its species. In other words, for the $n$th system particle ($n=1, 2$) and a bath particle belonging to $\alpha$th species ($\alpha=1, 2$), the stiffness $\textsf K_{n,\alpha}$  is given by $\textsf K_{n,\alpha} = k_{\rm I}$ for $n=\alpha$ and $\textsf K_{n,\alpha} = \kappa$ for $n\neq \alpha$. By setting $\tilde \Phi = 0 $, the interaction potential $V_{\rm I} = H_{\rm I}$ is then written as
\begin{equation}
V_{\rm I} (\boldsymbol{x},\tilde{\boldsymbol{x}}) = \frac{1}{2}
\sum_{n,\alpha,\tilde{n}_\alpha}
\textsf K_{n,\alpha}
\left ( x_n – \tilde x_{\tilde{n}_\alpha} \right)^2.
\end{equation}
Here, $\tilde n_{\alpha}  $ ($1\leq \tilde n_\alpha \leq \tilde N_{\alpha}$) is the index for a bath particle of the $\alpha$th species and $\tilde N_\alpha$ denotes the number of bath particles belonging to the $\alpha$th species. Here we consider the case $\tilde N_1 = \tilde N_2 \equiv \tilde N/2$. The nonzero off-diagonal stiffness $\kappa$ of the $\textsf K$ matrix renders the disjoint separability broken. Meanwhile, the translational invariance is maintained since all interactions are pairwise.

Similar to the first example, the potential of the mean force can be easily obtained from the rearranged form of the interaction potential as
\begin{equation}
V_{\rm I} ( \boldsymbol{x},\tilde{\boldsymbol{x}}) = 
\sum_{\alpha,\tilde{n}_{\alpha}} \frac{1}{2} (k_{\rm I} + \tilde k) \tilde X_{\tilde n_\alpha}^2 + \frac{\tilde N k_{\rm I} \kappa}{2( k_{\rm I} + \kappa )} (x_1-x_2)^2,
\end{equation}
where $\tilde X_{\tilde n_{\alpha}} = x_{\tilde n_{\alpha}} -\sum_n \textsf K_{n,\alpha} x_n /(k_{\rm I}+\kappa) $. The potential of the mean force is
\begin{equation}
	\Delta (\boldsymbol{x}) = \frac{1}{2} k (x_1-x_2)^2 + c, \label{eq:mean_force_ex2}
\end{equation}
where $k = \tilde{N} k_{\rm I} \kappa / 2( k_{\rm I} + \kappa )$ and $c$ is a constant without $\boldsymbol{x}$ dependence. Equation~\eqref{eq:mean_force_ex2} indicates that the two-system particles are coupled via an effective harmonic potential with stiffness $k$, which originates from the harmonic-interaction chain between $x_1$ and $x_2$ through the bath particles.

The effective damping tensor is also analytically solvable through the similar technique used for the first example. From Appendix~\ref{sec:app3}, it is 
\begin{equation}\label{eq:damping_ex2}
\textsf G = 
\frac{\tilde{\gamma} \tilde{N}}{2 (k_{\rm I} + \kappa)^2}
\begin{pmatrix}
k_{\rm I}^2 + \kappa^2 & 2 k_{\rm I} \kappa \\
2 k_{\rm I} \kappa & k_{\rm I}^2 + \kappa^2
\end{pmatrix}.
\end{equation}
$\Delta (\boldsymbol{x})$ and $\mathsf{G}$ are invariant under the permutation of $k_\textrm{I}$ and $\kappa$, as this permutation is equivalent to the exchange of the bath species. Thus, we can set $\kappa \leq k_\textrm{I}$ without loss of generality.
In the limit $\kappa \rightarrow 0$, where the disjoint separability is restored, $k$ in Eq.~\eqref{eq:mean_force_ex2} vanishes and $\textsf G $ in Eq.~\eqref{eq:damping_ex2} loses the information on the SB interaction potential and becomes the identity matrix with the overall factor consistent with Eq.~\eqref{eq:restored_G}. In the opposite limit, i.e. $\kappa = k_\textrm{I}$, the effective damping tensor becomes singular because the bath particles are coupled to only one normal mode $x_1 + x_2$.

To check whether the effective dynamics of this two-particle system follows Eq.~\eqref{eq:reduced_Langevin} with Eqs.~\eqref{eq:mean_force_ex2} and \eqref{eq:damping_ex2} numerically, we performed a simulation using the microscopic equations of motion~\eqref{eq:micro_bath}. The parameters are set to be the same as in the first example. 
To ensure the stability of the stationary state, we added an external harmonic force $- k_{\rm S} \boldsymbol{x}$ with stiffness $k_{\rm S} = 1.0$ to the system particles.

After the system reaching its equilibrium state, we evaluated the effective stiffness and damping tensor by means of measuring the variance of distance between the two particles and calculating the Green-Kubo formula, respectively. 
Figures~\ref{fig:ex2}(b) and (c) show the plots for $k/(\tilde N k_{\rm I})$ and $\textsf{G}_{n,m}/(\tilde N \tilde \gamma)$ against $\kappa/k_{\rm I}$ for various $k_{\rm I}$, respectively. All numerical data coincide with the analytic expectations. This clearly verifies that for a translationally invariant system the strong coupling effect emerges in the effective dynamics when the disjoint separability is broken. It also demonstrates the validity of our formalism in the presence of an external force applied to a system.

\subsection{Effect of interaction range on disjoint separability} \label{sec:separability}

\begin{figure}
	\centering
	\includegraphics[width = \columnwidth]{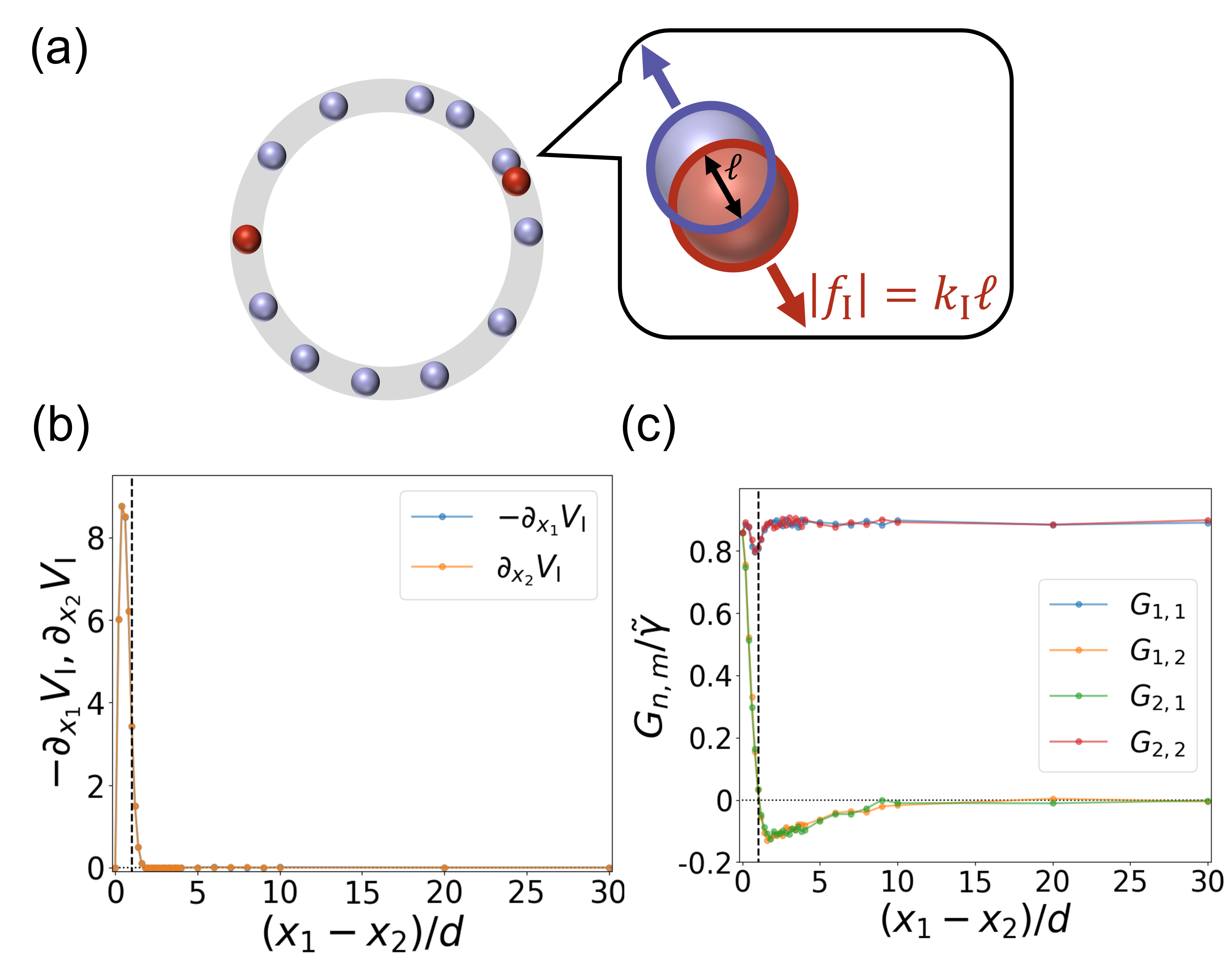}
	\caption{Results of the MD simulation.
		(a) Schematic of the two-particle system. Red and blue balls denote system and bath particles, respectively.
		$\ell = d- |x_n - \tilde{x}_{\tilde{n}}|$ represents the overlapped length between system and bath particles. Arrows indicate the soft repulsion forces whose strength is proportional to $\ell$.
		(b) Plot for mean force versus $(x_1 - x_2)/L$. Blue and orange curves represent the mean force applied to the first particle and minus of the mean force for the second particle, respectively.
		(c) Plot for $\textsf G_{m,n}/\tilde{\gamma}$ versus $(x_1 - x_2)/L$. Blue, orange, green, and red colors in (c) represent the data for $\textsf G_{1,1}$, $\textsf G_{1,2}$, $\textsf G_{2,1}$ and $\textsf G_{2,2}$, respectively.
        In both (b) and (c), the vertical dashed line indicates the interaction range $(x_1-x_2)/d=1$.
	}
	\label{fig:ex3}
\end{figure}


Different from the toy models illustrated in Secs.~\ref{sec:example1} and \ref{sec:example2}, in a usual experimental setup, where multiple particles of the system are immersed together in a single bath, it is almost impossible to satisfy the disjoint separability in a strict sense for several reasons. 
First, a bath particle can move around freely and thus its interacting partner among the system particles may change over time. Therefore,
members of the disjoint subgroup are not fixed in time. Second, a
bath particle can interact with several system particles simultaneously
when the range of the SB interaction is not sufficiently shorter than
the distance between system particles.

In this section, we numerically demonstrate that the disjoint separability can be effectively satisfied when the range of SB interaction is much shorter than the distance between system particles, even though the members of the disjoint subgroups change over time. For this purpose, we performed an MD simulation for a two-particle system moving on a one-dimensional ring with length $L$, as illustrated in Fig.~\ref{fig:ex3}(a). No external force is applied to the two particles and no interaction force exists between them ($\boldsymbol{f} = 0$). There also exist $\tilde N$ bath particles moving on the ring that are in contact with the thermostat. Similar to the system particles, no external potential is given to the bath particles, and no interaction force exists between them ($\tilde{\Phi} = 0$). The diameter of the system and bath particles is set to be the same as $d$. The interaction potential $U(x_n, \tilde x_{\tilde n})$ between a system particle at $x_n$ and a bath particle at $ \tilde x_{\tilde n}$ is as follows:
\begin{equation}
	U(x_n, \tilde x_{\tilde n}) = \frac{k_{\rm I}}{2} (x_n - \tilde x_{\tilde n})^2 - k_{\rm I} d |x_n - \tilde x_{\tilde n}|,
\end{equation}
when the system and bath particles overlap, that is, $|x_n - \tilde x_{\tilde n}| \leq d$. When there is no overlap, $U(x_n, \tilde x_{\tilde n})=0$. The total interaction potential is then given by $V_{\rm I} (\boldsymbol{x}, \tilde{\boldsymbol{x}}) = \sum_{n, \tilde n} U(x_n, \tilde x_{\tilde n})$. This form of repulsion potential is often employed in various MD simulations~\cite{han2021fluctuating}. The parameters of this simulation are set as $\tilde{N} = 10^3, \tilde{\gamma} = 10^{-2}, m = 10^{-2}, T = 10, k_{\rm I} = 10, d = 1$ and $L = 100$. To reduce the computation time, overdamped Langevin dynamics is assumed for the bath particles, i.e., $\tilde{m} = 0$.
After the whole system reaches equilibrium, the mean force can be evaluated using the relation $\boldsymbol{\nabla}_{x_n} \Delta(\boldsymbol{x}) = \langle \{\boldsymbol{\nabla}_{x_n} V_{\rm I}(\boldsymbol{x},\tilde{\boldsymbol{x}})\} \rangle_{\rm b}^\textrm{eq}$ from Eq.~\eqref{def:mean_force_potential} and the damping tensor $\textsf G_{n,m}$ can be obtained from Eq.~\eqref{def:damping_tensor}. Since the translational invariance holds, $\Delta $ and $\mathsf{G}$ satisfy Eq.~\eqref{eq:conditionI}, guaranteeing that they are functions of $x_1-x_2$ as explained in Appendix~\ref{sec:app4}.

Figures~\ref{fig:ex3}(b) and (c) show the dependence of $\Delta $ and $\mathsf{G}_{n,m} $ on the relative distance between the system particles $(x_1-x_2)/d$. In the region $x_1 - x_2 > 10 d$, where the range of SB interaction is much shorter than the distance between system particles, $\Delta $ and $\mathsf{G}_{n,m}$ exhibit no dependence of $x_1-x_2$. Especially, off-diagonal elements of $\textsf G_{n,m}$ $(n\neq m)$ vanish in this region. Therefore, the mean force term disappears and $\textsf G$ becomes a diagonal and constant matrix, which is consistent with the expectation achieved when disjoint separability is satisfied. On the other hand, for $x_1 - x_2 < 10 d$, where a bath particle can interact with multiple system particles simultaneously or members of the disjoint subgroup can change rather quickly, both $\Delta$ and $\mathsf{G}$ exhibit a certain dependence of $x_1-x_2$. Therefore, the mean force and off-diagonal elements of $\textsf G$ do not vanish, and thus, the dynamics relies on the SB interaction.

These observations highlight the dependence of disjoint separability's validity on the relative distance between system particles in real-world experiments. If the range of SB interaction is much shorter than the distance between system particles, disjoint separability holds effectively. This means that members of disjoint sets of bath particles might not be fixed in time, yet constituents of a disjoint subgroup can  remain stable for a considerable duration. Consequently, under such circumstances, the system's observed dynamics simply follows the conventional Langevin equation, without showing strong coupling effects. 
On the contrary, if the distance between system particles approaches the scale of SB interaction, the dynamics deviate from the conventional Langevin description. Detecting these SB coupling effects in a multi-particle experiment poses an intriguing and formidable challenge. Since these effects manifest within a restricted spatial domain, densely packing the system particles becomes essential to achieve a distance between them that is comparable to the SB interaction range. Moreover, a measurement apparatus with high spatial resolution is indispensable due to the typically minute scale of SB interaction compared to the system particle size. Without this meticulous experimental preparation, the system's behavior may appear to follow conventional Langevin dynamics, potentially masking the strong coupling effects.

\section{Discussion and Perspective}

It is instructive to compare our results with those of previous approaches used for deriving the conventional Langevin equation from a microscopic equation of motion. 
The first such approach uses the kinetic theory based on the Kramers–Moyal expansion introduced to study the Brownian motor~\cite{van2004microscopic, meurs2004rectification, lee2021geometry} and adiabatic piston~\cite{meurs2004rectification, piasecki1999adiabatic}. In these models, the system consists of a single degree of freedom, the SB interaction is given by the hard-core collision, and no other potential is applied to bath particles; accordingly, translational invariance and disjoint separability are satisfied. Furthermore, the bath particles are assumed to be always in equilibrium, which amounts to an infinitely fast equilibrating thermostat being attached to the bath particles. The microscopic setups of these models are the special case of our general setup satisfying the two conditions described in Sec.~\ref{sec:two_conditions}. Therefore, the conventional Langevin equation is derived when the limit of time-scale separation is taken into account. 

The second approach is the Caldeira–Leggett model~\cite{caldeira1983path}. The system of this model also consists of a single particle; thus, disjoint separability is satisfied. However, since the SB interaction and the potential applied to the bath particles are not in pairwise forms, the translational invariance is not satisfied. This may lead to a question why the mean force term does not appear in the resulting Langevin equation, even though the translational invariance of the potential is broken. This is due to the counter-term conventionally added to the total Hamiltonian of the Caldeira–Leggett model. In the derivation of the Langevin equation, the mean-force-like term is eventually canceled out by the counter-term. 
We also note that this approach does not take any explicit time-scale separation methods. Instead, the special form of interactions and spectral functions are assumed, which make the system dynamics Markovian.

We anticipate that our formalism will open the way to investigate thermodynamics for stochastic systems strongly coupled to baths and be utilized to simulate such systems without directly performing an MD simulation, so as to reduce the computational cost significantly.

\section*{Acknowledgements}

Authors acknowledge the Korea Institute for Advanced Study for providing computing resources 
(KIAS Center for Advanced Computation Linux Cluster System).
This research was supported by NRF Grants No.~2017R1D1A1B06035497 (H.P.), and individual KIAS Grants No.~QP013601 (H.P.), No.~PG064901 (J.S.L.) and No.~PG074002 (J.-M.P.) at the Korea Institute for Advanced Study.
This research was supported by an appointment to the JRG Program at the APCTP through the Science and Technology Promotion Fund and Lottery Fund of the Korean Government (J.-M.P.). This was also supported by the Korean Local Governments - Gyeongsangbuk-do Province and Pohang City (J.-M.P.).

\begin{appendices}

\section{Derivation of Mesoscopic SDE}\label{sec:app1}

The derivation of Eq.~\eqref{eq:reduced_Langevin} is based on a proper time-scale separation, where the bath variables are treated as fast variables, thus, integrated out under certain conditions. The derivation procedure can be divided into two steps. In the first step, the velocity variables of the bath are integrated out in the limit of small $\tilde m / \tilde \gamma $, which corresponds to the usual overdamped limit. Then, the corresponding Fokker-Plank equation reads
\begin{equation}
	\dot{P} (\boldsymbol{x},\boldsymbol{v},\tilde{\boldsymbol{x}},t)
	= \left ( \mathcal{L} + \frac{1}{\tilde \gamma }\tilde{\mathcal{L}}_{\rm o} \right)
	{P} (\boldsymbol{x},\boldsymbol{v},\tilde{\boldsymbol{x}},t), \label{eqM:FPeq}
\end{equation}
where $P (\boldsymbol{x},\boldsymbol{v},\tilde{\boldsymbol{x}},t)$ is probability distribution function for the whole system and the Fokker-Planck operators with respect to the system and the (overdamped) bath are given by
\begin{align}
	&\mathcal{L} = -\boldsymbol{\nabla}_x^{\textsf T} \boldsymbol{v} -\frac{1}{m} \boldsymbol{\nabla}_v^{\textsf T}
	\left[ \boldsymbol{f} (\boldsymbol{x},\boldsymbol{v},t)
	– \{ \boldsymbol{\nabla}_x V_\textrm{I}(\boldsymbol{x},\tilde{\boldsymbol{x}}) \} \right], \nonumber\\
	&\tilde{\mathcal{L}}_{\rm o} = 
	\boldsymbol{\nabla}_{\tilde{x}}^{\textsf T}
	\left[  \{\boldsymbol{\nabla}_{\tilde{x}} V_\textrm{I}(\boldsymbol{x},\tilde{\boldsymbol{x}}) \} + T 
	\boldsymbol{\nabla}_{\tilde{x}} \right], \label{eqM:operators}
\end{align}
respectively, where $\boldsymbol{\nabla}_v = (\partial_{v_1},  \cdots,\partial_{v_N})^{\textsf T}$.  The curly bracket $\{ {\scriptstyle{\square}} \}$ in Eq.~\eqref{eqM:operators} is used to indicate that derivative operators inside the bracket have no effect on terms outside of it. The overall factor $\tilde{\gamma}^{-1}$ in front of $\tilde{\mathcal{L}}_{\rm o}$ in Eq.~\eqref{eqM:FPeq} indicates that the relaxation time scale of the bath is proportional to the damping constant. Thus, the time separation can be implemented by taking the limit of small $\tilde{\gamma} $. This method is known as adiabatic elimination~\cite{risken1996fokker}. For completeness, we present the detailed procedure in the following.

We first consider the eigenfunctions $\varphi_k (\tilde{\boldsymbol{x}}|\boldsymbol{x})$ of $\tilde{\mathcal{L}}_{\rm o}$ and corresponding eigenvalues $-\lambda_k$ for a given $\boldsymbol{x}$, that is,
\begin{equation}
	\tilde{\mathcal{L}_{\rm o}} \varphi_k (\tilde{\boldsymbol{x}}| \boldsymbol{x})
	= - \lambda_k \varphi_k (\tilde{\boldsymbol{x}}| \boldsymbol{x}).
\end{equation}
Note that in our sign convention $\lambda_k$ is always positive except for $\lambda_0 = 0$ which is the eigenvalue of the stationary-state distribution $\varphi_0 (\tilde{\boldsymbol{x}}|\boldsymbol{x})$ of $\tilde{\mathcal{L}}_{\rm o}$ given by
\begin{equation}
	\varphi_0 (\tilde{\boldsymbol{x}}|\boldsymbol{x}) = \frac{1}{\mathcal Z_{\rm I} (\boldsymbol{x})} e^{- \beta V_{\rm I} (\boldsymbol{x}, \tilde{\boldsymbol{x}})}
\end{equation}
with the partition function $\mathcal Z_{\rm I} (\boldsymbol{x}) = \int d\tilde{\boldsymbol{x}} e^{-\beta V_{\rm I} (\boldsymbol{x},\tilde{\boldsymbol{x}})}$. Thus, $\varphi_0 (\tilde{\boldsymbol{x}}|\boldsymbol{x})$ is an equilibrium distribution of the overdamped bath for a given $\boldsymbol{x}$. 

Now we expand the probability distribution function in terms of $\varphi_k (\tilde{\boldsymbol{x}}|\boldsymbol{x})$ as
\begin{equation}
	P (\boldsymbol{x},\boldsymbol{v},\tilde{\boldsymbol{x}},t) = \sum_k C_k (\boldsymbol{x},\boldsymbol{v},t) \varphi_k (\tilde{\boldsymbol{x}}|\boldsymbol{x}). \label{eqM:pdf_expansion}
\end{equation}
We define $\varphi_k^\dagger (\tilde{\boldsymbol{x}}|\boldsymbol{x})$ as an eigenfunction of the adjoint operator $\tilde{\mathcal{L}}_{\rm o}^\dagger$. Then it satisfies the eigenvalue equation $\tilde{\mathcal{L}}^\dagger \varphi_k^\dagger (\tilde{\boldsymbol{x}}|\boldsymbol{x}) = -\lambda_k \varphi_k^\dagger (\tilde{\boldsymbol{x}}|\boldsymbol{x})$ and the orthogonality $\int d\tilde{\boldsymbol{x}} \varphi_k^\dagger (\tilde{\boldsymbol{x}}|\boldsymbol{x}) \varphi_m (\tilde{\boldsymbol{x}}|\boldsymbol{x}) = \delta_{k,m}$. By multiplying $\varphi_k^\dagger (\tilde{\boldsymbol{x}}|\boldsymbol{x})$ to both sides of Eq.~\eqref{eqM:pdf_expansion}, integrating over $\tilde{\boldsymbol{x}}$, and using the orthogonality, we arrive at the following equation for the coefficient $C_k (\boldsymbol{x},\boldsymbol{v},t)$ as
\begin{equation}
	\dot{C}_k (\boldsymbol{x},\boldsymbol{v},t)
	= \sum_{m\geq 0}  \mathcal{F}_{k,m} C_m (\boldsymbol{x},\boldsymbol{v},t) 
	- \frac{\lambda_k}{\tilde \gamma} C_k (\boldsymbol{x},\boldsymbol{v},t),  \label{eqM::C_k_eq}
\end{equation}
where the tilted-system-Fokker-Planck operator $\mathcal{F}_{k,m}$ is defined as
\begin{equation}\label{def:tilted_op}
	\mathcal{F}_{k,m} \equiv \int d\tilde{\boldsymbol{x}} \varphi_k^\dagger(\tilde{\boldsymbol{x}}|\boldsymbol{x})
	\mathcal{L} \varphi_m(\tilde{\boldsymbol{x}}|\boldsymbol{x}).
\end{equation}
In the small $\tilde{\gamma}$ limit, one can write Eq.~\eqref{eqM::C_k_eq} up to the linear order in $\tilde \gamma$ as
\begin{align}
	&\dot{C}_0 (\boldsymbol{x},\boldsymbol{v},t) = 
	\sum_{m \geq 0}
	\mathcal{F}_{0,m} C_m (\boldsymbol{x},\boldsymbol{v},t)~~~~~~~~~~~~{\rm for}~k=0 , \nonumber \\
	&C_k(\boldsymbol{x},\boldsymbol{v},t) = \frac{\tilde \gamma}{\lambda_k}
	\mathcal{F}_{k,0} C_0 (\boldsymbol{x},\boldsymbol{v},t) + \mathcal{O} (\tilde{\gamma}^2)~~~~{\rm for~}k\geq 1 . \label{eqM:C_k_eq}
\end{align}
By inserting the expression for $C_k$ into the equation for $C_0$ in Eq.~\eqref{eqM:C_k_eq}, we obtain the uncoupled equation of motion for  $C_0$ as follows: 
\begin{equation}
	\dot{C}_0 (\boldsymbol{x},\boldsymbol{v},t) = \mathcal{L}_{\rm r} C_0 (\boldsymbol{x},\boldsymbol{v},t) + \mathcal{O} (\tilde{\gamma}^2),  \label{eqM:reducedLangevin}
\end{equation}
where the reduced-system operator $\mathcal L_{\rm r}$ is 
\begin{equation}
	\mathcal{L}_{\rm r} \equiv \mathcal{F}_{0,0} + \tilde \gamma \sum_{k \geq 1}
	\frac{\mathcal{F}_{0,k} \mathcal{F}_{k,0}}{\lambda_k}.  \label{eqM:reduced_FP_operator}
\end{equation}
Using $\varphi_0^\dagger (\tilde{\boldsymbol{x}}|\boldsymbol{x})=1$, we can show that $C_0$ is the marginal distribution of $P (\boldsymbol{x},\boldsymbol{v},\tilde{\boldsymbol{x}},t)$ as
\begin{equation}
	C_0 (\boldsymbol{x},\boldsymbol{v},t) = \int d\tilde{\boldsymbol{x}} \varphi_0^\dagger (\tilde{\boldsymbol{x}}|\boldsymbol{x}) P (\boldsymbol{x},\boldsymbol{v},\tilde{\boldsymbol{x}},t) = P(\boldsymbol{x},\boldsymbol{v},t)~.
\end{equation}
Thus, Eq.~\eqref{eqM:reducedLangevin} is the Fokker-Planck equation for the reduced system.

The explicit form of $\mathcal{L}_{\rm r}$ can be obtained from the definition of the tilted operators in Eq.~\eqref{def:tilted_op}. By using the orthogonality of the eigenfunctions and the mean-force relation $ \boldsymbol{\nabla}_x \Delta(\boldsymbol{x}) =  \int d\tilde{\boldsymbol{x}}
\{\boldsymbol{\nabla}_x V_{\rm I}(\boldsymbol{x},\tilde{\boldsymbol{x}})\}
\varphi_0(\tilde{\boldsymbol{x}}|\boldsymbol{x}) $ from Eq.~\eqref{def:mean_force_potential}, one can show that
\begin{equation}
	\mathcal{F}_{0,0} = 
	- \boldsymbol{\nabla}_x^{\textsf T} \boldsymbol{v}
	- \frac{1}{m} \boldsymbol{\nabla}_v^{\textsf T}
	\left[
	\boldsymbol{f} (\boldsymbol{x},\boldsymbol{v},t)
	- \{\boldsymbol{\nabla}_x \Delta(\boldsymbol{x}) \} \label{eqM:F_00}
	\right].
\end{equation}
Therefore, $\mathcal{F}_{0,0}$ corresponds to the deterministic evolution part of the reduced equation of motion in Eq.~\eqref{eq:reduced_Langevin}. 
Using the eigenfunction orthogonality and $\varphi_0^\dagger (\tilde{\boldsymbol{x}}|\boldsymbol{x})=1$, it is straightforward to show that
\begin{align}
	\mathcal{F}_{0,k} = \frac{1}{m}
	\boldsymbol{\nabla}_v^{\textsf T}
	\boldsymbol{b}_k \label{eqM:F_0k}
\end{align}
with $\boldsymbol{b}_{k}\equiv \int d\tilde{\boldsymbol{x}} \{\boldsymbol{\nabla}_x V_{\rm I} (\boldsymbol{x},\tilde{\boldsymbol{x}}) \} \varphi_k(\tilde{\boldsymbol{x}}|\boldsymbol{x})$.
Similarly, we can also show that 
\begin{align}
	\mathcal{F}_{k,0} =
	\left (
	\frac{1}{T} \boldsymbol{v} + \frac{1}{m} \boldsymbol{\nabla}_v
	\right )^{\textsf T} \boldsymbol{b}_k. \label{eqM:F_k0}
\end{align}
For deriving Eq.~\eqref{eqM:F_k0}, the relations $\boldsymbol{\nabla}_x^{\textsf T} \boldsymbol{v} \varphi_0(\tilde{\boldsymbol{x}}|\boldsymbol{x}) =  \boldsymbol{v}^{\textsf T} \{ \boldsymbol{\nabla}_x  \varphi_0(\tilde{\boldsymbol{x}}|\boldsymbol{x}) \} +  \boldsymbol{v}^{\textsf T} \varphi_0(\tilde{\boldsymbol{x}}|\boldsymbol{x}) \boldsymbol{\nabla}_x$,
$\boldsymbol{\nabla}_x \varphi_0(\tilde{\boldsymbol{x}}|\boldsymbol{x}) = -\beta \{ \boldsymbol{\nabla}_x V_\textrm{I}  \} \varphi_0(\tilde{\boldsymbol{x}}|\boldsymbol{x}) - \{\boldsymbol{\nabla}_x \ln \mathcal Z_{\rm I} \} \varphi_0(\tilde{\boldsymbol{x}}|\boldsymbol{x})$, and $\varphi_k(\tilde{\boldsymbol{x}}|\boldsymbol{x})=\varphi_k^\dagger(\tilde{\boldsymbol{x}}|\boldsymbol{x}) \varphi_0(\tilde{\boldsymbol{x}}|\boldsymbol{x})$ are used in order.  Plugging Eqs.~\eqref{eqM:F_00}, \eqref{eqM:F_0k}, and \eqref{eqM:F_k0} into Eq.~\eqref{eqM:reduced_FP_operator} yields  
\begin{align}
	\mathcal{L}_{\rm r}=
	&
	-\boldsymbol{\nabla}_x^{\textsf T} \boldsymbol{v}
	- \frac{1}{m}
	\boldsymbol{\nabla}_v^{\textsf T}
	\left[ \boldsymbol{f} (\boldsymbol{x}, \boldsymbol{v}, t)
	- \{\boldsymbol{\nabla}_x \Delta(\boldsymbol{x}) \}
	\right]    +\frac{1}{m}
	\boldsymbol{\nabla}_v^{\textsf T}
	\mathsf{G} (\boldsymbol{x})
	\left(\boldsymbol{v}+\frac{T}{m}\boldsymbol{\nabla}_v\right)
\end{align}
with the damping tensor
\begin{equation}\label{def:damping_tensor_alt}
	\mathsf{G} (\boldsymbol{x})
	\equiv
	\frac{\tilde \gamma}{T}\sum_{k\geq1}\frac{\boldsymbol{b}_k \boldsymbol{b}_k^{\rm T}}{\lambda_k} .
\end{equation}
From Eq.~\eqref{def:damping_tensor_alt} it is clear that the damping tensor is symmetric.

The remaining task is showing that Eq.~\eqref{def:damping_tensor_alt}  is equivalent to Eq.~\eqref{def:damping_tensor}. From the definition of $\boldsymbol{b}_k$, we rewrite the damping tensor as
\begin{align}\label{eqM:G_1}
	\mathsf{G} (\boldsymbol{x})
	&= \frac{\tilde \gamma}{T} \int d\tilde{\boldsymbol{x}} \int d{\tilde{\boldsymbol{x}}}^\prime
	\left\{\boldsymbol{\nabla}_x V_{\rm I} \left(\boldsymbol{x},\tilde{\boldsymbol{x}}\right)\right\}
	\{ \boldsymbol{\nabla}_x^{\textsf T} V_{\rm I} \left({\boldsymbol{x},\tilde{\boldsymbol{x}}}^\prime\right) \}	
	\sum_{k \geq 1} \frac{\varphi_k(\tilde{\boldsymbol{x}}|\boldsymbol{x}) \varphi_k^\dagger (\tilde{\boldsymbol{x}}^\prime|\boldsymbol{x})}{\lambda_k} 
	\varphi_0 (\tilde{\boldsymbol{x}}^\prime|\boldsymbol{x}) \nonumber \\
	&= \frac{\tilde \gamma}{T} \int d\tilde{\boldsymbol{x}} \int d{\tilde{\boldsymbol{x}}}^\prime
	\left\{\boldsymbol{\nabla}_x V_{\rm I} \left(\boldsymbol{x},\tilde{\boldsymbol{x}}\right)\right\}
	\{ \boldsymbol{\nabla}_x^{\textsf T} V_{\rm I} \left({\boldsymbol{x},\tilde{\boldsymbol{x}}}^\prime\right) \}	
	\left\{ -\tilde{\mathcal L}_{\rm o}^+ \delta(\tilde{\boldsymbol{x}} - \tilde{\boldsymbol{x}}^\prime)	\right\}
	\varphi_0 (\tilde{\boldsymbol{x}}^\prime|\boldsymbol{x}), 
\end{align}
where the Moore-Penrose inverse ${\tilde{\mathcal{L}}}_{\rm o}^+$ of ${\tilde{\mathcal{L}}}_{\rm o}$ satisfies
\begin{equation} \label{eqM:MP_inverse}
	{\tilde{\mathcal{L}}}_{\rm o}^+ \left(\tilde{\boldsymbol{x}}\right) h\left(\tilde{\boldsymbol{x}}\right)
	= -\sum_{k\geq1}
	{\frac{\varphi_k\left(\tilde{\boldsymbol{x}}|\boldsymbol{x}\right)}{\lambda_k} \int{d\tilde{\boldsymbol{x}}'' \varphi_k^\dag\left(\tilde{\boldsymbol{x}}''|\boldsymbol{x}\right)
			h\left(\tilde{\boldsymbol{x}}''\right)}}
\end{equation}
for an arbitrary function $h\left(\tilde{\boldsymbol{x}}\right)$.
Then, ${\tilde{\mathcal{L}}}_{\rm o}^+ \delta\left(\tilde{\boldsymbol{x}}-{\tilde{\boldsymbol{x}}}^\prime\right) $ can be manipulated as
\begin{align} \label{eqM:sum_k}
	{\tilde{\mathcal{L}}}_{\rm o}^+ \delta\left(\tilde{\boldsymbol{x}}-{\tilde{\boldsymbol{x}}}^\prime\right) 
	&= -\sum_{k \geq 1} \frac{\varphi_k(\tilde{\boldsymbol{x}}|\boldsymbol{x}) \varphi_k^\dagger (\tilde{\boldsymbol{x}}^\prime|\boldsymbol{x})}{\lambda_k}  \nonumber \\
	&= -\frac{1}{\tilde \gamma}\int_0^\infty dt \sum_{k \geq 1} e^{-\frac{\lambda_k}{\tilde \gamma} t} \varphi_k(\tilde{\boldsymbol{x}}|\boldsymbol{x}) \varphi_k^\dagger (\tilde{\boldsymbol{x}}^\prime|\boldsymbol{x})
	\nonumber \\
	&= -\frac{1}{\tilde \gamma} \int_0^\infty dt  \left[ e^{\frac{\tilde{\mathcal L}_{\rm o}}{\tilde \gamma} t}\sum_{k \geq 0}  \varphi_k(\tilde{\boldsymbol{x}}|\boldsymbol{x}) \varphi_k^\dagger (\tilde{\boldsymbol{x}}^\prime|\boldsymbol{x}) - \varphi_0 (\tilde{\boldsymbol{x}}|\boldsymbol{x})\right]
	\nonumber \\
	&= -\frac{1}{\tilde \gamma} \int_0^\infty dt  \left[ P(\tilde{\boldsymbol{x}},t|\tilde{\boldsymbol{x}}^\prime,0)  - \varphi_0 (\tilde{\boldsymbol{x}}|\boldsymbol{x})\right],
\end{align}
where $P(\tilde{\boldsymbol{x}},t|\tilde{\boldsymbol{x}}^\prime,0) = e^{\tilde{\mathcal L}_{\rm o} t} \delta (\tilde{\boldsymbol{x}} - \tilde{\boldsymbol{x}}^\prime) $ is a propagator starting from $\tilde{\boldsymbol{x}}^\prime$ at time $0$ and reaching  $\tilde{\boldsymbol{x}}$ at time $t$. For deriving Eq.~\eqref{eqM:sum_k}, $\varphi_0^\dagger (\tilde{\boldsymbol{x}}|\boldsymbol{x}) = 1$ and $\sum_{k \geq 0}  \varphi_k(\tilde{\boldsymbol{x}}|\boldsymbol{x}) \varphi_k^\dagger (\tilde{\boldsymbol{x}}^\prime|\boldsymbol{x}) = \delta (\tilde{\boldsymbol{x}} - \tilde{\boldsymbol{x}}^\prime)$ are used for the third and the last equalities, respectively. 
By plugging Eq.~\eqref{eqM:sum_k} into Eq.~\eqref{eqM:G_1}, we finally arrive at
\begin{align} \label{eqM:G_tenson_final}
	\mathsf{G} (\boldsymbol{x})
	&= \frac{1}{T}  \int_0^\infty dt \left[ \int d\tilde{\boldsymbol{x}} \int d{\tilde{\boldsymbol{x}}}^\prime
	\left\{\boldsymbol{\nabla}_x V_{\rm I} \left(\boldsymbol{x},\tilde{\boldsymbol{x}}\right)\right\}
	\{ \boldsymbol{\nabla}_x^{\textsf T} V_{\rm I} \left({\boldsymbol{x},\tilde{\boldsymbol{x}}}^\prime\right) \}	
	P(\tilde{\boldsymbol{x}},t|\tilde{\boldsymbol{x}}^\prime,0)\varphi_0 (\tilde{\boldsymbol{x}}^\prime|\boldsymbol{x}) \right. \nonumber \\
	&~~~~~~~~~~~~~~~~~~~~~~~~~~~~~~~~~~~~~~ ~~~~~~~~~~~~~~~~~~~~~~
	\left. - \langle \boldsymbol{\nabla}_x V_{\rm I} \left(\boldsymbol{x},\tilde{\boldsymbol{x}}\right) \rangle_{\rm b}^{\rm eq}  \langle \boldsymbol{\nabla}_x^{\textsf T} V_{\rm I} \left(\boldsymbol{x},\tilde{\boldsymbol{x}}\right) \rangle_{\rm b}^{\rm eq} \right]  \nonumber\\
	&=  \frac{1 }{T}  \int_0^\infty dt \left[ \langle \{ \boldsymbol{\nabla}_x V_{\rm I} \left(\boldsymbol{x},\tilde{\boldsymbol{x}}(t)\right) \}  \{ \boldsymbol{\nabla}_x^{\textsf T} V_{\rm I} \left(\boldsymbol{x},\tilde{\boldsymbol{x}}(0)\right) \} \rangle_{\rm b}^{\rm eq} 
	- \langle \boldsymbol{\nabla}_x V_{\rm I} \left(\boldsymbol{x},\tilde{\boldsymbol{x}}\right) \rangle_{\rm b}^{\rm eq}  \langle \boldsymbol{\nabla}_x^{\textsf T} V_{\rm I} \left(\boldsymbol{x},\tilde{\boldsymbol{x}}\right) \rangle_{\rm b}^{\rm eq}
	\right] \nonumber \\
	&= \frac{1}{T} \int_0^\infty dt~
	C_{\boldsymbol{\nabla}_x V_\textrm{I}, \boldsymbol{\nabla}_x^{\textsf T} V_\textrm{I}} (t|\boldsymbol{x}),  
\end{align}
where $\langle \cdots \rangle_{\rm b}^{\rm eq}$ is an average with respect to the bath-equilibrium distribution $\varphi_0 (\tilde{\boldsymbol{x}}|\boldsymbol{x})$ for a given $\boldsymbol{x}$.

\section{Equation of motion for the center of mass coordinate of the bath particles} \label{sec:method2}

For speeding up the MD simulation, we used the center of mass coordinates of the bath particles. For the first example in Sec.~\ref{sec:example1}, these coordinates are given by $\tilde{X}\left(t\right)=\sum_{\tilde{n}}{{\tilde{x}}_{\tilde{n}}\left(t\right)}/\tilde{N}$ and $\tilde{V}\left(t\right)=\sum_{\tilde{n}}{{\tilde{v}}_{\tilde{n}}\left(t\right)}/\tilde{N}$ for the position and velocity of the bath particles, respectively. Then, the dynamics of the total system can be described in the reduced space, which is governed by the following two equations of motion:
\begin{align} \label{eqM:reducedEq1}
m{\dot{v}}_1\left(t\right)
&=-\tilde{N}k_{\rm I} \left(x_1\left(t\right)-\tilde{X}\left(t\right)\right), \nonumber \\
\tilde{m}\tilde{V}\left(t\right)
&=-k_{\rm I} \left(\tilde{X}\left(t\right)-x_1\left(t\right)\right)-\tilde{k}\tilde{X}\left(t\right)-\tilde{\gamma}\tilde{V}\left(t\right)+\tilde{\xi}\left(t\right)
\end{align}
with $\tilde{\xi}\left(t\right) \equiv \sum_{\tilde{n}}{{\tilde{\xi}}_{\tilde{n}}\left(t\right)}/\tilde{N}$. Equation~\eqref{eqM:reducedEq1} enables us to calculate the system dynamics with significantly reduced computational cost for large $\tilde N$ case.

Similarly, for the second example in Sec.~\ref{sec:example2}, we consider the center of mass coordinates of the bath particles of the same species, defined as
\begin{equation}
{\tilde{X}}_\alpha\left(t\right)\equiv\frac{2}{\tilde{N}}\sum_{\tilde{n}_\alpha}{{\tilde{x}}_{\tilde{n}_\alpha}\left(t\right)},~~ {\tilde{V}}_\alpha\left(t\right)\equiv\frac{2}{\tilde{N}}\sum_{\tilde{n}_\alpha}{{\tilde{v}}_{\tilde{n}_\alpha}\left(t\right)}.
\end{equation}
Then, the equations of motion in the reduced space are written as
\begin{align}
m{\dot{v}}_n\left(t\right)
&=-k_{\rm S} x_n\left(t\right) -\frac{\tilde{N}}{2} \sum_{\alpha} \textsf K_{n,\alpha} \left ( x_n\left(t\right) -{\tilde{X}}_\alpha\left(t\right)\right ) , \\
\tilde{m}{\dot{V}}_\alpha\left(t\right)
&=-\sum_{n}{\textsf K_{\alpha,n} \left({\tilde{X}}_\alpha \left(t\right)- x_n\left(t\right)\right)}-\tilde{\gamma}{\tilde{V}}_\alpha\left(t\right)+{\tilde{\xi}}_\alpha\left(t\right),
\end{align}
where $-k_{\rm S} {\boldsymbol{x}}$ is an external force added for confining the system and ${\tilde{\xi}}_\alpha\left(t\right) \equiv 2\sum_{\tilde{n}_\alpha}{{\tilde{\xi}}_{\tilde{n}_\alpha}\left(t\right)}/\tilde{N}$.

\section{Pairwise property of translationally invariant function}
\label{sec:app4}

Suppose $f(x_1, x_2, \cdots, x_N)$ is a translationally invariant function satisfying $f(x_1 + c, x_2 +c, \cdots, x_N+c) = f(x_1, x_2, \cdots, x_N)$ for an arbitrary constant $c$. For a set of constants $\{ c_i\}$,  where $i= 1, 2, \cdots, N$, we can write
\begin{align}
	f(x_1, x_2, \cdots, x_N) = \frac{1}{N} \sum_{i=1}^N f(x_1 + c_i, x_2 +c_i, \cdots, x_N+c_i).
\end{align}
By taking $c_n = -x_n$, the function can be written as
\begin{align}
	f(x_1, x_2, \cdots, x_N) = \frac{1}{N} \sum_{i=1}^N f(x_1 - x_i, x_2  - x_i, \cdots, x_N - x_i).
\end{align}
Thus, a translationally invariant function can be always expressed as  a function of difference of all pairs of $x_i$.

\section{Generalized Green-Kubo relation}
\label{sec:app2}

Here, we revisit the generalized Green-Kubo (GK) relation~\cite{pavliotis2015stochastic} and derive Eq.~\eqref{eq:GK_rel} by showing that it is a special case of the generalized GK relation. We consider a Langevin dynamics governed by a Langevin operator $L$. The system state and the steady-state distribution are denoted by $\boldsymbol{y}$ and $\varphi_0 (\boldsymbol{y})$, respectively. One can define the generalized drift coefficient $V^h\left(\boldsymbol{y}\right)$ and diffusion coefficient $D^{h,g}\left(\boldsymbol{y}\right)$ as
\begin{align}
& V^h\left(\boldsymbol{y}\right) = \lim_{dt\rightarrow0}{\frac{1}{dt}}\left\langle h\left(\boldsymbol{y}+d \boldsymbol{y} \right)-h\left(\boldsymbol{y}\right)\right\rangle, \nonumber \\
& D^{h,g}\left(\boldsymbol{y}\right) = \lim_{dt\rightarrow0}
{\frac{\left\langle\left(h\left(\boldsymbol{y}+d\boldsymbol{y}\right)-h\left(\boldsymbol{y}\right)\right)\left(g\left(\boldsymbol{y}+d\boldsymbol{y}\right)-g\left(\boldsymbol{y}\right)\right)\right\rangle}{dt}},
\end{align}
where $h\left(\boldsymbol{y}\right)$ and $g\left(\boldsymbol{y}\right)$ are arbitrary functions. The generalized GK relation reads
\begin{equation}
\left\langle D^{h,g}\left(\boldsymbol{y}\right)\right\rangle_s \equiv \int d \boldsymbol{y} \varphi_0 (\boldsymbol{y}) D^{h,g}(\boldsymbol{y}) =\int_{0}^{\infty} dt K_{V^h,V^g}(t)~,
\end{equation}
where $\left\langle \cdots \right\rangle_s$ stands for the average at the stationary state and $K_{V^h,V^g}\left(t\right) = C_{V^h,V^g}\left(t\right) + C_{V^g,V^h}\left(t\right)$ is a symmetrized stationary state correlation. Here, the function $C_{V^h,V^g} (t)$ is the same one defined in Eq.~\eqref{def:correl_func} without specification of $\boldsymbol{x}$. 

To derive the generalized GK relation, it is convenient to rewrite the generalized coefficients in terms of the Langevin operator as follows. First, the drift coefficient is reformulated as
\begin{align} \label{eqA:Vh_expression}
V^h \left(\boldsymbol{y}\right) &=  \lim_{dt\rightarrow0}{\frac{1}{dt}} \int d\boldsymbol{y}'  \left\{   h(\boldsymbol{y}')- h(\boldsymbol{y}) \right\} P(\boldsymbol{y}',t+dt|\boldsymbol{y},t)
\nonumber \\
&= \int d\boldsymbol{y}'  \left\{   h(\boldsymbol{y}')- h(\boldsymbol{y}) \right\} \lim_{dt\rightarrow0} \frac{P(\boldsymbol{y}',t+dt|\boldsymbol{y},t) - \delta(\boldsymbol{y}' - \boldsymbol{y})}{dt} 
\nonumber \\
&= \int d\boldsymbol{y}'  h(\boldsymbol{y}')   L(\boldsymbol{y}') \delta(\boldsymbol{y}' - \boldsymbol{y})  - \int d\boldsymbol{y}'   h(\boldsymbol{y}) \lim_{t' \rightarrow t} \frac{d}{dt} P(\boldsymbol{y}',t'|\boldsymbol{y},t)
\nonumber \\
&= \int d\boldsymbol{y}'  \delta(\boldsymbol{y}' - \boldsymbol{y}) L^\dagger (\boldsymbol{y}')   h(\boldsymbol{y}')  - h(\boldsymbol{y})  \lim_{t' \rightarrow t} \frac{d}{dt} \int d\boldsymbol{y}'   P(\boldsymbol{y}',t'|\boldsymbol{y},t)
\nonumber \\
&= L^\dag (\boldsymbol{y}) h (\boldsymbol{y}). 
\end{align}
$\int d\boldsymbol{y}'  \left\{   h(\boldsymbol{y}')- h(\boldsymbol{y}) \right\} \delta(\boldsymbol{y}' - \boldsymbol{y}) = 0$ and
$\int d\boldsymbol{y}'   P(\boldsymbol{y}',t'|\boldsymbol{y},t) = 1$ are used for deriving the second and the last equalities of Eq.~\eqref{eqA:Vh_expression}, respectively. Similarly, we can manipulate the diffusion coefficient as
\begin{align}  \label{eqA:Dhg_expression}
D^{h,g}(\boldsymbol{y}) &=   \int d\boldsymbol{y}'  \left\{   h(\boldsymbol{y}')- h(\boldsymbol{y}) \right\} \left\{  g(\boldsymbol{y}')- g(\boldsymbol{y}) \right\} \lim_{dt\rightarrow0} \frac{P(\boldsymbol{y}',t+dt|\boldsymbol{y},t) - \delta(\boldsymbol{y}' - \boldsymbol{y})}{dt}
\nonumber \\
&= \int d\boldsymbol{y}'  \left\{ h(\boldsymbol{y}') g(\boldsymbol{y}') - h(\boldsymbol{y}')g(\boldsymbol{y})- h(\boldsymbol{y})g(\boldsymbol{y}')  \right\} L(\boldsymbol{y}') \delta(\boldsymbol{y}' - \boldsymbol{y})  \nonumber \\
&~~~~~~~~~~~~~~~~~~~~~~~~~~~~~~~~~~~~~~~~~~~~~~~
-  h(\boldsymbol{y})  g(\boldsymbol{y})  \lim_{t' \rightarrow t} \frac{d}{dt}  \int d\boldsymbol{y}'   P(\boldsymbol{y}',t'|\boldsymbol{y},t)
\nonumber \\
&= \int d\boldsymbol{y}'  \delta(\boldsymbol{y}' - \boldsymbol{y}) \left\{ L^\dagger (\boldsymbol{y}')   h(\boldsymbol{y}')  g(\boldsymbol{y}') -  g(\boldsymbol{y}) L^\dagger (\boldsymbol{y}')   h(\boldsymbol{y}')  -  h(\boldsymbol{y}) L^\dagger (\boldsymbol{y}')    g(\boldsymbol{y}') \right\} 
\nonumber \\
&= L^\dag (\boldsymbol{y}) h(\boldsymbol{y}) g(\boldsymbol{y})-h(\boldsymbol{y}) L^\dag (\boldsymbol{y}) g(\boldsymbol{y})-g(\boldsymbol{y}) L^\dag (\boldsymbol{y}) h(\boldsymbol{y}).
\end{align}

From Eq.~\eqref{eqM:G_tenson_final}, we can write
\begin{align} \label{eqA:C_hg_expression}
\int_0^\infty dt C_{V^h , V^g} (t)
&=   \int d \boldsymbol{y} \int d \boldsymbol{y}^\prime
V^h \left(\boldsymbol{y}\right)
V^g ( \boldsymbol{y}^\prime ) 
\int_0^\infty dt \left\{ P(\boldsymbol{y},t|\boldsymbol{y}^\prime,0)  - \varphi_0 (\boldsymbol{y})	\right\} \varphi_0 (\boldsymbol{y}^\prime) \nonumber \\
&= \int d \boldsymbol{y} \int d \boldsymbol{y}^\prime
\left\{ L^\dagger (\boldsymbol{y}) h\left(\boldsymbol{y}\right) \right\}
\left\{ L^\dagger (\boldsymbol{y}^\prime) g( \boldsymbol{y}^\prime ) \right\} 
\left\{ -L^+ (\boldsymbol{y}) \delta\left(\boldsymbol{y}-\boldsymbol{y}^\prime\right)  \right\} \varphi_0 (\boldsymbol{y}^\prime) \nonumber \\
&=  -\int d \boldsymbol{y} \int d \boldsymbol{y}^\prime \delta\left(\boldsymbol{y}-\boldsymbol{y}^\prime\right) 
\left\{ L^\dagger (\boldsymbol{y}^\prime) g( \boldsymbol{y}^\prime ) \right\}  \varphi_0 (\boldsymbol{y}^\prime)
\left\{  (L^\dagger)^+ (\boldsymbol{y}) L^\dagger (\boldsymbol{y}) h\left(\boldsymbol{y}\right) \right\} ,
\end{align}
where Eqs.~\eqref{eqM:sum_k}, \eqref{eqA:Vh_expression}, and \eqref{eqA:Dhg_expression} are used for the second equality in Eq.~\eqref{eqA:C_hg_expression}. For the last equality in Eq.~\eqref{eqA:C_hg_expression}, $(L^+)^\dagger (\boldsymbol{y}) = (L^\dagger)^+ (\boldsymbol{y})$ is used since
\begin{align}
\int d \boldsymbol{y} h(\boldsymbol{y}) (L^+)^\dagger g(\boldsymbol{y}) 
&= \int d \boldsymbol{y} g(\boldsymbol{y})  L^+ h(\boldsymbol{y}) \nonumber \\
&= - \sum_{k\geq1} \frac{1}{\lambda_k} \int d \boldsymbol{y} g(\boldsymbol{y})  
\varphi_k ( \boldsymbol{y}) \int d \boldsymbol{y}' \varphi_k^\dag (\boldsymbol{y}') h\left(\boldsymbol{y}'\right) 
\nonumber \\
&= \int d \boldsymbol{y}'  h\left(\boldsymbol{y}'\right) \left\{- \sum_{k\geq1} \frac{\varphi_k^\dag (\boldsymbol{y}')}{\lambda_k} \int d \boldsymbol{y} g(\boldsymbol{y})  
\varphi_k ( \boldsymbol{y})  \right\}
\nonumber \\
&= \int d \boldsymbol{y} h(\boldsymbol{y})  (L^\dag)^+ g(\boldsymbol{y}),
\end{align} 
where $\varphi_k$ and $\varphi_k^\dagger$ are the $k$th eigenfunctions of $L$ and $L^\dagger$, respectively, and Eq.~\eqref{eqM:MP_inverse} is used for the second and the final equalities. 

To proceed further, we need one more relation:
\begin{align}\label{eq:prod_MP_inv_ori}
(L^\dag)^+ L^\dag h(\boldsymbol{y}) 
&= (L^\dag)^+ L^\dag \sum_{k\geq 0} \varphi_k^\dagger (\boldsymbol{y}) \int{d \boldsymbol{y}^\prime}
\varphi_k (\boldsymbol{y}^\prime)
h (\boldsymbol{y}^\prime)
\nonumber \\
&= -(L^\dag)^+ \sum_{k\geq 0} \lambda_k \varphi_k^\dagger (\boldsymbol{y}) \int{d \boldsymbol{y}^\prime}
\varphi_k (\boldsymbol{y}^\prime)
h (\boldsymbol{y}^\prime)
\nonumber \\
&=  \sum_{k\geq 1} \varphi_k^\dagger (\boldsymbol{y}) \int{d \boldsymbol{y}^\prime}
\varphi_k (\boldsymbol{y}^\prime)
h (\boldsymbol{y}^\prime)
\nonumber \\
&=  h (\boldsymbol{y}) - \int{d \boldsymbol{y}^\prime}
\varphi_0 (\boldsymbol{y}^\prime)
h (\boldsymbol{y}^\prime),
\end{align}
where the expansion of $h(\boldsymbol{y})$ in terms of the eigenfunctions, $h(\boldsymbol{y}) = \sum_{k\geq 0} \varphi_k^\dagger (\boldsymbol{y}) \int{d \boldsymbol{y}^\prime}
\varphi_k (\boldsymbol{y}^\prime)
h (\boldsymbol{y}^\prime)$, is used for the first and the final equalities and Eq.~\eqref{eqM:MP_inverse} is used for the third equality in Eq.~\eqref{eq:prod_MP_inv_ori}. Plugging Eq.~\eqref{eq:prod_MP_inv_ori} into Eq.~\eqref{eqA:C_hg_expression}, we
obtain
\begin{align} \label{eqA:C_hg_final}
\int_{0}^{\infty}dt C_{V^h,V^g}\left(t\right)
&= - \int d \boldsymbol{y}\varphi_0 (\boldsymbol{y}) h(\boldsymbol{y})  L^\dagger g(\boldsymbol{y}) + \int d \boldsymbol{y} \varphi_0 (\boldsymbol{y}) L^\dagger g(\boldsymbol{y}) \int d \boldsymbol{y}^\prime \varphi_0 (\boldsymbol{y}^\prime) h(\boldsymbol{y}^\prime).
\end{align}
The second term in the right hand side of Eq.~\eqref{eqA:C_hg_final} vanishes since $\int d \boldsymbol{y} \varphi_0 (\boldsymbol{y}) L^\dagger g(\boldsymbol{y}) = \int d \boldsymbol{y} g(\boldsymbol{y}) L \varphi_0 (\boldsymbol{y})  = 0$. Similarly, $\int d \boldsymbol{y} \varphi_0 (\boldsymbol{y}) \{ L^\dag h(\boldsymbol{x}) g(\boldsymbol{x}) \}= 0$. Therefore, using Eqs.~\eqref{eqA:C_hg_final} and \eqref{eqA:Dhg_expression}, we finally arrive at 
\begin{align}
&\int_{0}^{\infty} dt  K_{V^h,V^g}\left(t\right)
= \int_{0}^{\infty}dt C_{V^h,V^g}\left(t\right) + \int_{0}^{\infty}dt C_{V^g,V^h}\left(t\right)
\nonumber \\
&=
\int d \boldsymbol{y} \varphi_0 (\boldsymbol{y}) \{ L^\dag h(\boldsymbol{y}) g(\boldsymbol{y}) \}
- \int d \boldsymbol{y} \varphi_0 (\boldsymbol{y}) h(\boldsymbol{y})  L^\dagger g(\boldsymbol{y})
- \int d \boldsymbol{y} \varphi_0 (\boldsymbol{y}) g(\boldsymbol{y})  L^\dagger h(\boldsymbol{y}) 
\nonumber \\
&= \int d \boldsymbol{y} \varphi_0 (\boldsymbol{y}) D^{h,g}(\boldsymbol{y}).
\end{align}

For a specific case where $h\left(\tilde{\boldsymbol{x}}\right)={\tilde{x}}_{\tilde{n}}$ and $g\left(\tilde{\boldsymbol{x}}\right)={\tilde{x}}_{\tilde{m}}$ for our model in the main text, the generalized coefficients become
\begin{equation}
V^h\left(\tilde{\boldsymbol{x}}\right)=
-\frac{1}{\tilde{\gamma}}\partial_{\tilde{x}_{\tilde{n}}}V_{\rm I}\left(\boldsymbol{x},\tilde{\boldsymbol{x}}\right),\ \ V^g\left(\tilde{\boldsymbol{x}}\right)=
-\frac{1}{\tilde{\gamma}}\partial_{{\tilde{x}}_{\tilde{m}}}V_{\rm I}\left(\boldsymbol{x},\tilde{\boldsymbol{x}}\right)
\end{equation}
and
\begin{equation}
D^{h,g}\left(\tilde{\boldsymbol{x}}\right)=\frac{2T}{\tilde{\gamma}}\delta_{\tilde{n},\tilde{m}}~.
\end{equation}
Thus, Eq.~\eqref{eq:GK_rel} is a consequence of the generalized GK relation with a special choice of $h$ and $g$ functions.

\section{Derivation of Eqs.~\eqref{eq:damping_ex1} and \eqref{eq:damping_ex2}}
\label{sec:app3}

For both examples illustrated in Figs.~\ref{fig:ex1} and \ref{fig:ex2}, the interaction potentials can be decomposed as
\begin{equation}\label{eq:VI_decomposition}
V_{\rm I} \left(\boldsymbol{x},\tilde{\boldsymbol{x}}\right)
= \sum_{\tilde{n}}{V_{{\rm I},\tilde{n}}\left(\boldsymbol{x},\tilde{x}_{\tilde{n}}\right)}
\end{equation}
since no direct interaction between bath particles exists. Therefore, the Fokker-Planck equation can be written in terms of the each bath-particle segment  as
\begin{equation}
{\tilde{\mathcal{L}}}_{\rm o}  = \sum_{\tilde{n}} {\tilde{\mathcal{L}}}_{{\rm o},\tilde{n}}, 
\end{equation}
where ${\tilde{\mathcal{L}}}_{{\rm o},\tilde{n}}$ is defined by
\begin{equation}
{\tilde{\mathcal{L}}}_{{\rm o},\tilde{n}} 
\equiv \frac{1}{\tilde{\gamma}}
\partial_{\tilde{x}_{\tilde{n}}}
\left ( \left \{\partial_{\tilde{x}_{\tilde{n}}}
V_{{\rm I},\tilde{n}}\left(\boldsymbol{x},{\tilde{x}}_{\tilde{n}}\right)\right \}
+T\partial_{\tilde{x}_{\tilde{n}}} \right ).
\end{equation}
Then, eigenfunctions and eigenvalues of ${\tilde{\mathcal{L}}}_{{\rm o}}$ are given by
\begin{equation}\label{eq:phi_decomposition}
\varphi_{\boldsymbol{k}}\left(\tilde{\boldsymbol{x}} | \boldsymbol{x}\right)
=\prod_{\tilde{n}}{\varphi_{k_{\tilde{n}}}
	\left({\tilde{x}}_{\tilde{n}} |  \boldsymbol{x}\right)}
~\textrm{ and }~
\lambda_{\boldsymbol{k}}=\sum_{\tilde{n}}\lambda_{k_{\tilde{n}}},
\end{equation}
respectively, where $\boldsymbol{k}=(k_1,\ldots,k_{\tilde{N}})$, $\varphi_{k_{\tilde{n}}}({\tilde{x}}_{\tilde{n}}| \boldsymbol{x})$ is the eigenfuction of ${\tilde{\mathcal{L}}}_{{\rm o},\tilde{n}} $, and 
$\lambda_{k_{\tilde{n}}}$ is the eigenvalue of $\varphi_{k_{\tilde{n}}} ({\tilde{x}}_{\tilde{n}}| \boldsymbol{x})$.
Using Eqs.~\eqref{eq:VI_decomposition}, \eqref{eq:phi_decomposition}, and the orthogonality
$\int{d\tilde{x}_{\tilde n}}\varphi_{k_{\tilde{n}}} (\tilde{x}_{\tilde{n}}| \boldsymbol{x}) = \int{d\tilde{x}_{\tilde n}} \varphi_0^\dag (\tilde{x}_{\tilde n}| \boldsymbol{x}) \varphi_{k_{\tilde{n}}} (\tilde{x}_{\tilde n} | \boldsymbol{x}) = \delta_{k_{\tilde{n}},0}$,
we can rewrite Eq.~\eqref{def:damping_tensor_alt} as
\begin{equation}\label{eq:G_decomposition}
\mathsf{G}(\boldsymbol{x}) = \frac{1}{T}
\sum_{\tilde{n}=1}^{\tilde{N}}\sum_{k_{\tilde{n}}\geq1}
\frac{\boldsymbol{b}_{k_{\tilde{n}}}
	\boldsymbol{b}^{\textsf T}_{k_{\tilde{n}}}}{\lambda_{k_{\tilde{n}}}}
\end{equation}
with $\boldsymbol{b}_{k_{\tilde{n}}} = \int{d{\tilde{x}}_{\tilde{n}}}
\left\{\boldsymbol{\nabla}_x V_{{\rm I},\tilde{n}}\left(\boldsymbol{x},{\tilde{x}}_{\tilde{n}}\right)\right\}
\varphi_{k_{\tilde{n}}}\left({\tilde{x}}_{\tilde{n}} | \boldsymbol{x}\right)$.

Now we focus on the first example, that is, the single-particle model without translational invariance. 
Since every $V_{{\rm I},\tilde{n}}\left(x_1,{\tilde{x}}_{\tilde{n}}\right) \equiv \hat{V}_{\rm I} (x_1,\tilde{x}_{\tilde{n}})$ is identical for each bath particle, Eq.~\eqref{eq:G_decomposition} is reduced to 
\begin{equation}\label{eq:damping_alt_ex1}
\gamma=\frac{\tilde{N}}{T}\sum_{k\geq1}\frac{b_k^2}{\lambda_k},
\end{equation}
where $b_k = -k_{\rm I} \int{d{\tilde{x}}_1} \tilde{x}_1 \varphi_k\left(\tilde{x}_1 | x_1\right)$ and $\lambda_k$ is the eigenvalue of the Fokker-Planck operator $\tilde{\mathcal{L}}_{{\rm o},1}$ of a single-bath particle. Thus, from now on, it is sufficient to consider the case of the 1st bath particle, i.e., $\tilde n = 1$.  Since the interaction potentials are harmonic, it is convenient to use the Hermitianized Fokker-Planck operator defined by
\begin{equation}
{\tilde{\mathcal{H}}}_{{\rm o},1} \equiv\frac{1}{\sqrt{\varphi_0 \left(\tilde{x}_1 | x_1\right)}}{\tilde{\mathcal{L}}}_{{\rm o},1} \sqrt{\varphi_0\left(\tilde{x}_1 | x_1 \right)}
= - \frac{k_I + \tilde{k}}{\tilde{\gamma}}{\tilde{a}}_+{\tilde{a}}_-
\end{equation}
with the ladder operators given by
\begin{align} \label{eqA:ladder_op_ex1}
{\tilde{a}}_+&\equiv\sqrt{\frac{k_{\rm I}+\tilde{k}}{4T}}
\left(\tilde{x}_1-\frac{k_{\rm I}}{k_{\rm I} +\tilde{k}}x_1\right)
-\sqrt{\frac{T}{k_{\rm I}+\tilde{k}}}\partial_{\tilde{x}_1}, 
\nonumber \\
{\tilde{a}}_-&\equiv\sqrt{\frac{k_{\rm I}+\tilde{k}}{4T}}
\left(\tilde{x}_1-\frac{k_{\rm I}}{k_{\rm I}+\tilde{k}}x_1\right)
+\sqrt{\frac{T}{k_{\rm I}+\tilde{k}}}\partial_{\tilde{x}_1}.
\end{align}
We define $\psi_k\left({\tilde{x}}_1 |x_1\right)$ as the eigenfunction of $\tilde{\mathcal{H}}_{{\rm o},1}$. Then, as usually done in a quantum harmonic oscillator, the operators satisfy $\left[{\tilde{a}}_-,{\tilde{a}}_+\right] =1$,
${\tilde{a}}_- \psi_k\left({\tilde{x}}_1|x_1\right)=\sqrt k\psi_{k-1}\left({\tilde{x}}_1 |x_1 \right)$,  and
${\tilde{a}}_+\psi_k\left({\tilde{x}}_1 |x_1\right)=\sqrt{k+1}\psi_{k+1}\left({\tilde{x}}_1 |x_1\right)$
with eigenvalue of $\psi_k\left({\tilde{x}}_1 |x_1\right)$
\begin{equation}\label{eq:eigenvalue_ex1}
\lambda_k=\frac{k_{\rm I} + \tilde{k}}{\tilde{\gamma}} k.
\end{equation}
Using the relation
$\psi_k\left({\tilde{x}}_1|x_1\right) =\left(\sqrt{\varphi_0\left(\tilde{x}_1 | x_1\right)}\right)^{-1}\varphi_k\left({\tilde{x}}_1 |x_1 \right)=\sqrt{\varphi_0\left({\tilde{x}}_1 |x_1\right)}\varphi_k^\dag\left({\tilde{x}}_1 |x_1\right)$,
we can rewrite $b_k$ as
\begin{equation} \label{eqA:bk_ex1}
b_k = -k_{\rm I} \int{dx_1}\psi_0\left(\tilde{x}_1 | x_1\right)
\tilde{x}_1
\psi_k\left(\tilde{x}_1 | x_1\right)~.
\end{equation}
From Eq.~\eqref{eqA:ladder_op_ex1}, we have
\begin{equation} \label{eqA:x_1_ex1}
\tilde{x}_1 =
\sqrt{\frac{T}{k_{\rm I} + \tilde{k}}}
\left ( \tilde{a}_+ + \tilde{a}_- \right )
+ \frac{k_{\rm I}}{k_{\rm I} + \tilde{k}} x_1.
\end{equation}
Plugging Eq.~\eqref{eqA:x_1_ex1} into Eq.~\eqref{eqA:bk_ex1}, we obtain 
\begin{equation} \label{eqA:bk1_ex1}
b_k=-k_{\rm I} \sqrt{\frac{T}{k_{\rm I}+\tilde{k}}}\delta_{k,1} ~~~{\rm for}~k\geq 1.
\end{equation}
Using Eqs.~\eqref{eq:damping_alt_ex1}, \eqref{eq:eigenvalue_ex1}, and \eqref{eqA:bk1_ex1}, we finally arrive at Eq.~\eqref{eq:damping_ex1}.

Next, we consider the two-particle model without disjoint separability. In this model, we can decompose the interaction potential as
\begin{align}
&V_{\rm I} (\boldsymbol{x}, \tilde{\boldsymbol{x}}) 
=\sum_{\tilde{n}}
\left\{ 
\hat{V}_{\rm I} \left(x_1,x_2,{\tilde{x}}_{\tilde{n}_1} \right) +
\hat{V}_{\rm I} \left(x_2,x_1,{\tilde{x}}_{\tilde{n}_2}\right)
\right\},  
\end{align}
where $\hat{V}_{\rm I}\left(x_1,x_2,\tilde{x}\right)=\frac{1}{2}k_{\rm I} \left(x_1-\tilde{x}\right)^2+\frac{1}{2}\kappa\left(x_2-\tilde{x}\right)^2$. Then, Eq.~\eqref{eq:G_decomposition} is reduced  to
\begin{equation}
\mathsf{G} =
\frac{\widetilde{N}}{2T} \sum_{k\geq1}
\left (
\frac{\boldsymbol{b}_k\boldsymbol{b}_k^{\textsf T}}{\lambda_k}
+
\frac{\bar{\boldsymbol{b}}_k\bar{\boldsymbol{b}}_k^{\textsf T}}{\lambda_k}
\right ),
\end{equation}
where $\boldsymbol{b}_k$ and $\bar{\boldsymbol{b}}_k$ are given by
\begin{align}
\boldsymbol{b}_k &\equiv \int{d{\tilde{x}}}
\left\{\boldsymbol{\nabla}_x \hat{V}_{\rm I}
\left(x_1,x_2,{\tilde{x}}\right)\right\}
\varphi_k\left({\tilde{x}} | \boldsymbol{x}\right), \nonumber \\
\bar{\boldsymbol{b}}_k &\equiv
\int{d{\tilde{x}}}
\left\{\boldsymbol{\nabla}_x \hat{V}_{\rm I}
\left(x_2,x_1,{\tilde{x}}\right)\right\}
\varphi_k\left({\tilde{x}} | \boldsymbol{x}\right)
=
\begin{pmatrix}
	0 & 1 \\ 1 & 0
\end{pmatrix}
\boldsymbol{b}_k.
\end{align}
Similar to the first example, it is convenient to use the Hermitianized Fokker-Planck operator for $\tilde{\mathcal{L}}_{{\rm o},1}$ with the potential $\hat{V}_{\rm I} \left(x_1,x_2,\tilde{x}\right)$ as follows:
\begin{equation}
{\tilde{\mathcal{H}}}_{\rm o,1}
= - \frac{k_{\rm I} + \kappa}{\tilde{\gamma}}{\tilde{a}}_+{\tilde{a}}_-
\end{equation}
with the ladder operators
\begin{align} \label{eqA:ladder_oper2}
\tilde{a}_+&\equiv
\sqrt{\frac{k_{\rm I} + \kappa}{4 T}}
\left(\tilde{x}-\frac{k_{\rm I} x_1+\kappa x_2}{k_{\rm I}+\kappa}\right)
- \sqrt{\frac{T}{k_{\rm I} + \kappa}} \partial_{\tilde{x}}~,
\nonumber \\
\tilde{a}_-&\equiv
\sqrt{\frac{k_{\rm I} + \kappa}{4 T}}
\left(\tilde{x}-\frac{k_{\rm I} x_1+\kappa x_2}{k_{\rm I}+\kappa}\right)
+ \sqrt{\frac{T}{k_{\rm I} + \kappa}} \partial_{\tilde{x}}~.
\end{align}
Thus, the eigenvalue of ${\tilde{\mathcal{H}}}_{\rm o,1}$ is 
\begin{equation}
\lambda_k = \frac{k_{\rm I} + \kappa}{\tilde{\gamma}} k.
\end{equation}
Using Eq.~\eqref{eqA:ladder_oper2}, one can show that
\begin{align}
\boldsymbol{\nabla}_x \hat{V}_{\rm I} ( x_1, x_2, \tilde{x} )
=
- \sqrt{\frac{T}{k_{\rm I} + \kappa}}
(\tilde{a}_+ + \tilde{a}_-)
\begin{pmatrix}
	k_{\rm I} \\ \kappa
\end{pmatrix}
+
\frac{k_{\rm I} \kappa}{k_{\rm I} + \kappa}
\begin{pmatrix}
	x_1 - x_2 \\ x_2 - x_1
\end{pmatrix} , 
\end{align}
which leads to
\begin{equation} \label{eqA:bk_ex2}
\boldsymbol{b}_k
= - \sqrt{\frac{T}{k_{\rm I} +\kappa}} \delta_{k,1}
\begin{pmatrix}
	k_{\rm I} \\ \kappa 
\end{pmatrix} ~~~{\rm for}~k\geq1.
\end{equation}
Plugging Eq.~\eqref{eqA:bk_ex2} into Eq.~\eqref{eq:G_decomposition} and using $\lambda_1=(k_{\rm I}+\kappa)/T$ lead to Eq.~\eqref{eq:damping_ex2}.




\end{appendices}



\end{document}